%% file: main.tex
\documentclass[10pt,a4paper]{article}
\usepackage[a4paper, left=2cm, right=2cm, top=2cm]{geometry}

\input{./tex/settings.tex}
\input{./tex/plotting.tex}

\usetikzlibrary{external}
\title{Adjoint Node-Based Shape Optimization of Free Floating Vessels}

\begin{document}

\author[1]{Niklas K{\"u}hl\thanks{niklas.kuehl@tuhh.de}}
\author[2]{Thanh Tung Nguyen}
\author[2]{Michael Palm}
\author[2]{Dirk J{\"u}rgens}
\author[1]{Thomas Rung}

\affil[1]{Hamburg University of Technology, Institute for Fluid Dynamics and Ship Theory, Am Schwarzenberg-Campus 4, D-21075 Hamburg, Germany}
\affil[2]{J.M. Voith SE \& Co. KG, Alexanderstraße 2, D-89522 Heidenheim, Germany}

\providetoggle{tikzExternal}
%\settoggle{tikzExternal}{true}
\settoggle{tikzExternal}{false}

\maketitle

%%%%%%%%%%%%%%%%%%%%%%%%%%%%%%%%%%%%%%%%%%%%%%%%%%%%%%%%%%%%%%%%%%%%%%%%%%%%%%%
%
\begin{abstract}

The paper is concerned with a node-based, gradient-driven, continuous adjoint two-phase flow procedure to optimize the shapes of free-floating vessels and discusses three topics.
First, we aim to convey that elements of a Cahn-Hilliard formulation should augment the frequently employed Volume-of-Fluid two-phase flow model to maintain dual consistency. It is seen that such consistency serves as the basis for a robust primal/adjoint coupling in practical applications at huge Reynolds and Froude numbers.
The second topic covers different adjoint coupling strategies. A central aspect of the application is the floating position, particularly the trim and the sinkage, that interact with a variation of hydrodynamic loads induced by the shape updates. Other topics addressed refer to the required level of density coupling and a more straightforward --yet non-frozen-- adjoint treatment of turbulence. The third part discusses the computation of a descent direction within a node-based environment. We will illustrate means to deform both the volume mesh and the hull shape simultaneously and at the same time obey technical constraints on the vessel's displacement and its extensions.
The Hilbert-space approach provides smooth shape updates using the established coding infrastructure of a computational fluid dynamics algorithm and provides access to managing additional technical constraints.  
Verification and validation follow from a submerged 2D cylinder case. The application includes a full-scale offshore supply vessel at $\mathrm{Re} = 3 \cdot 10^8$ and $\mathrm{Fn} = 0.37$. Results illustrate that the fully parallel procedure can automatically reduce the drag of an already pre-optimized shape by 9--13\% within $\approx \mathcal{O}$(\SI{}{10.000}-\SI{}{30.000}) CPUh depending on the considered couplings and floatation aspects.

\end{abstract}
%%%%%%%%%%%%%%%%%%%%%%%%%%%%%%%%%%%%%%%%%%%%%%%%%%%%%%%%%%%%%%%%%%%%%%%%%%%%%%%
\section{Introduction}
\input{tex/introduction}

%%%%%%%%%%%%%%%%%%%%%%%%%%%%%%%%%%%%%%%%%%%%%%%%%%%%%%%%%%%%%%%%%%%%%%%%%%%%%%%
%
%%%%%%%%%%%%%%%%%%%%%%%%%%%%%%%%%%%%%%%%%%%%%%%%%%%%%%%%%%%%%%%%%%%%%%%%%%%%%%%
\section{Mathematical Model}
\label{sec:mamodel}
\input{tex/mathematical_model}

%%%%%%%%%%%%%%%%%%%%%%%%%%%%%%%%%%%%%%%%%%%%%%%%%%%%%%%%%%%%%%%%%%%%%%%%%%%%%%%
%
%%%%%%%%%%%%%%%%%%%%%%%%%%%%%%%%%%%%%%%%%%%%%%%%%%%%%%%%%%%%%%%%%%%%%%%%%%%%%%%
\section{Shape and Grid Update}
\label{sec:gridupdate}
\input{tex/grid_modifications}

%%%%%%%%%%%%%%%%%%%%%%%%%%%%%%%%%%%%%%%%%%%%%%%%%%%%%%%%%%%%%%%%%%%%%%%%%%%%%%%
%
%%%%%%%%%%%%%%%%%%%%%%%%%%%%%%%%%%%%%%%%%%%%%%%%%%%%%%%%%%%%%%%%%%%%%%%%%%%%%%%
\section{Validation}
\label{sec:validation}
\input{tex/validation}

%%%%%%%%%%%%%%%%%%%%%%%%%%%%%%%%%%%%%%%%%%%%%%%%%%%%%%%%%%%%%%%%%%%%%%%%%%%%%%%
%
%%%%%%%%%%%%%%%%%%%%%%%%%%%%%%%%%%%%%%%%%%%%%%%%%%%%%%%%%%%%%%%%%%%%%%%%%%%%%%%
\section{Application}
\label{sec:application}
\input{tex/application}

%%%%%%%%%%%%%%%%%%%%%%%%%%%%%%%%%%%%%%%%%%%%%%%%%%%%%%%%%%%%%%%%%%%%%%%%%%%%%%%
%
%%%%%%%%%%%%%%%%%%%%%%%%%%%%%%%%%%%%%%%%%%%%%%%%%%%%%%%%%%%%%%%%%%%%%%%%%%%%%%%
\section{Conclusion}
\label{sec:conclusion}

\input{tex/conclusion}

\section{Acknowledgements}
The current work was a part of the research projects "Dynamic Adaptation of Modular Shape Optimization Processes" funded by the German Federal Ministry for Economic Affairs and Energy (BMWi, Grant No. 03SX453B) and "Drag Optimization of Ship Shapes" funded by the German Research Foundation (DFG, Grant No. RU 1575/3-1). The research took place within the Research Training Group (RTG) 2583 "Modeling, Simulation, and Optimization with Fluid Dynamic Applications" funded by the German Research Foundation. The authors gratefully acknowledge this support. Selected computations were performed with resources provided by the North-German Super-computing Alliance (HLRN).
%%%%%%%%%%%%%%%%%%%%%%%%%%%%%%%%%%%%%%%%%%%%%%%%%%%%%%%%%%%%%%%%%%%%%%%%%%%%%%%
%
%%%%%%%%%%%%%%%%%%%%%%%%%%%%%%%%%%%%%%%%%%%%%%%%%%%%%%%%%%%%%%%%%%%%%%%%%%%%%%%

\end{document}

%% file: tex/settings.tex
\usepackage[hidelinks]{hyperref}

\usepackage{authblk}

\usepackage[nointlimits]{amsmath}
\usepackage{amsfonts}
\usepackage{siunitx}
\usepackage{mathtools}
\usepackage{array}

\usepackage{subfigure}
\usepackage{xcolor}
\usepackage{lscape}
\usepackage{pdfpages}

\usepackage[ruled,vlined]{algorithm2e}
\usepackage{etoolbox}
\usepackage{ulem}

\definecolor{mycolor_blue}{RGB}{66,124,161}
\definecolor{mycolor_grey}{RGB}{198,198,198}

\usepackage{doi}
\usepackage{natbib}

%% file: tex/plotting.tex
\usepackage{graphicx}
\usepackage{tikz}
\usepackage{pgfplots}

\usetikzlibrary{decorations.pathmorphing,arrows,intersections,arrows.meta,angles,quotes}

\pgfplotsset{
	kurze Legende/.style={
		legend image code/.code={
			\draw[##1,mark repeat=2,line width=0.6pt]
			plot coordinates {
				(0cm,0cm)
				(0.3cm,0cm)
			};
		}
	}
}

\pgfplotsset{
	compat = newest,
	scale only axis, 
	max space between ticks = 50pt,
	ticklabel style = {font=\footnotesize},
	legend style =  {font=\footnotesize},
	grid = major,
	grid style = {dotted},
	legend columns=1, 
	xtick pos=left,
	ytick pos=left
}

\pgfplotsset{select coords between index/.style 2 args={
		x filter/.code={
			\ifnum\coordindex<#1\fi
			\ifnum\coordindex>#2\fi
		}
}}

\definecolor{color1}{HTML}{0060AD} % blau
\definecolor{color2}{HTML}{FF4500} % rot
\definecolor{color3}{HTML}{FFA500} % gelb
\definecolor{color4}{HTML}{006400} % gruen
\definecolor{color5}{HTML}{9400D3} % lila
\definecolor{color6}{HTML}{800000} % bordeauxrot/braun
\definecolor{color7}{HTML}{000000} % schwarz
\definecolor{color8}{HTML}{0000FF} % blau heller
\definecolor{color9}{HTML}{FF0000} % rot heller
\definecolor{mycolor_blue}{RGB}{66,124,161}% niklas blue
\definecolor{mycolor_grey}{RGB}{198,198,198} % niklas grey

\tikzstyle{line1} = [color=color7,semithick] 
\tikzstyle{line2} = [color=color2,densely dotted,semithick]
\tikzstyle{line3} = [color=color1,densely dashed,semithick]
\tikzstyle{line4} = [color=color5,dash dot,semithick]
\tikzstyle{line5} = [color=color4,dash dot dot,semithick]
\tikzstyle{line6} = [color=color6,semithick]

\tikzstyle{mark1} = [color=color7,mark=x,mark size=2pt,mark options=solid,semithick] 
\tikzstyle{mark2} = [color=color2,mark=o,mark size=2pt,mark options=solid,semithick]
\tikzstyle{mark3} = [color=color1,mark=*,mark size=2pt,mark options=solid,semithick]
\tikzstyle{mark4} = [color=color5,mark=triangle,mark size=2pt,mark options=solid,semithick]
\tikzstyle{mark5} = [color=color4,mark=square,mark size=2pt,mark options=solid,semithick]
\tikzstyle{mark6} = [color=color7,mark=o,mark size=2pt,mark options=solid,semithick]
\tikzstyle{mark7} = [color=color7,mark=*,mark size=2pt,mark options=solid,semithick]
\tikzstyle{mark8} = [color=color7,mark=triangle,mark size=2pt,mark options=solid,semithick]

%% file: tex/introduction.tex
International shipping is responsible for transporting around 90\% of the global trade. The dominant role of shipping is attributable to the low fuel consumption per tonne-km of transported cargo. However, the mere magnitude of the many ten thousand operating vessels puts environmental and economic aspects of shipping into the focus of regulatory provisions. The seaborne pollution and approximately 50\% of the direct operating costs for shipping are related to fuel consumption, which in turn is governed by the vessel's resistance. Therefore, reducing the hydrodynamic drag, even by a few per mille, is highly appreciated from commercial and environmental perspectives. The endeavor for shortening development cycle times shifts the center of interest towards simulation-based approaches.

Marine engineering two-phase flow simulations mainly refer to Volume-of-Fluid (VoF) methods, cf. \cite{hirt1981volume}, which reconstruct the free surface from an indicator function that quantifies the volume concentration of the participating phases. The popularity of VoF methods is due to the simplicity of a shared kinematics approach, the inherently conservative formulation, and the capability to predict merging and rupturing of free surfaces. When attention is directed to the simulation-driven optimization of non-parameterized industrial shapes, gradient-based local optimization procedures using adjoint formulations are perhaps the most efficient approach to complement the simulation process by an optimization component. The efficiency benefits of the adjoint method increase with the number of degrees of freedom, and the procedure requires an established reference/initial design. Both aspects apply to the optimization of industrial shapes.  

Industrial applications of adjoint methods to optimize fluid dynamic shapes have reached an impressive level of maturity for single-phase flows, cf. \cite{othmer2014adjoint} or \cite{papoutsis2016continuous}. However, adjoint applications to marine engineering two-phase flows remain in their infancy. Significant challenges refer to the substantial Reynolds number turbulent flow and the immiscible two-phase flow characteristics that feature a discontinuous property change across the interface. Moreover, the dynamic floatation alters the drag, which interacts with the shape modification. In addition, hull shape updates are usually constraint to a sufficient level of smoothness and must conserve the vessel's displacement. Therefore, only a few applications were previously published for adjoint optimizations in marine engineering two-phase flows, cf. \cite{palacios2012shape, springer2015adjoint, kroger2018adjoint, he2019design}.
 
The adjoint analysis aims at the efficient computation of derivative information for an integral objective functional with respect to (w.r.t) a general control function, cf.  \cite{giles1997adjoint, giles2000introduction, kroger2018adjoint, papoutsis2019multi}. Two competing methods, known as the continuous and the discrete adjoint approach, are widely employed, cf.  \cite{peter2010numerical}. In continuous space, the dual or adjoint flow state can be interpreted as a co-state that follows from the primal flow model. However, the formulation of boundary conditions and the choice of an appropriate discretization of the underlying Partial Differential Equation (PDE) system is not intuitively obvious in a continuous adjoint framework. The situation gets more delicate for complex flow models with possibly non-differentiable expressions and larger PDE-systems featuring an augmented level of non-linearities. Related marine engineering examples refer to the aforementioned discontinuous property changes and the many inter-parameter couplings between the momentum/continuity equations on the one hand and the equation governing the indicator function on the other hand. Another frequently debated topic is the adjoint treatment of turbulence usually modeled by transport equations in a Reynolds-averaged Navier-Stokes (RANS) framework. Practical solutions found in the literature 
often suggest the neglect of adjoint variables and are frequently labeled ''incomplete'' or ''frozen'' adjoint strategies, such as the ''frozen turbulence'' or the ''frozen free surface'' approach, e.g. \cite{soto2004adjoint, dwight2006effects, martinelli2007adjoint, othmer2008continuous, stuck2012adjoint, marta2013handling, kroger2018adjoint}

Without a doubt, frozen adjoint strategies impair the computed sensitivity derivative and, at the same time, grossly simplify its calculation. The degree of derivative uncertainty is, however, debatable, cf. \cite{zymaris2010adjoint, hartmann2011adjoint, marta2013handling, papoutsis2015continuous, kavvadias2015continuous, manservisi2016numerical, manservisi2016optimal}. To assure consistent and synchronized primal and dual development states, discrete adjoint approaches based upon automatic differentiation were suggested by, e.g., \cite{nielsen2004implicit, nielsen2010discrete} or \cite{nielsen2013discrete, burghardt2022discrete}. The approach passes over the adjoint PDE and directly bridges a discrete linearized primal system into a consistent dual system, cf. \cite{giles1997adjoint, giles2000introduction} or \cite{vassberg2006aerodynamic, vassberg2006aerodynamic_II}. 

Despite the various merits and drawbacks of the discrete vs. the continuous adjoint method, the authors believe that the latter offers significant cost benefits for large-scale parallel implementations. Moreover, it is unique for its invaluable contribution to a physical understanding, i.e., the challenges mentioned above often disclose the weaknesses of the flow model. Nonetheless, considering the full range of inter-parameter couplings of a consistent framework can hamper the robustness and efficiency, mainly if a sequential or partly sequential algorithm is employed, while accuracy implications of a frozen adjoint approach are not well understood.

Therefore, the present contribution scrutinizes selected continuous adjoint formulations using a conventional pressure-based, sequential finite-volume algorithm, see \cite{ferziger2012computational}, for optimizing the shape of free-floating ships. Emphasis is given to two-phase flow $k-\omega$-type RANS procedures coupled to a motion modeler. A CAD-free shape update is used under restrictions of the length and the hull's displacement. Novel aspects refer to the adjoint two-phase flow approach and the application to high Reynolds number free-floating full-scale configurations using a simplified algebraic adjoint turbulence treatment. Moreover, we present a novel strategy to preserve the displacement and restrict the length within a volume-based identification of the descent direction. The employed formulation aims to balance accuracy opportunities and efficiency weaknesses using ''improved frozen'' approaches that retain the algorithmic benefits and preserve the predictive realism in practical marine flows.

The remainder is organized as follows: Section \ref{sec:mamodel} is concerned with the derivation of the mathematical model. Section \ref{sec:gridupdate} outlines our approach to adjust the floatation and to simultaneously update the mesh and the shape under the aegis of geometric constraints. Subsequently, the numerical method and the optimization procedure are briefly outlined in section \ref{sec:Numerik}. The \ref{sec:validation}th section is devoted to the validation for  2D flow around a submerged cylinder. Section \ref{sec:application} scrutinizes the performance of the optimizer for a total drag objective applied to an offshore supply vessel at full scale. The final section \ref{sec:conclusion} provides conclusions and outlines future research. The publication employs Einstein's summation convention for lower-case Latin subscripts. Vectors and tensors are defined with reference to Cartesian spatial coordinates, e.g., $x_k$, and the spatial derivative vector refers to $\nabla _k$.

%% file: tex/mathematical_model.tex
\subsection{Two-Phase Model}
The paper deals with the flow of two immiscible, inert fluids  ($a,b$) featuring constant bulk densities ($\rho_\mathrm{a}, \rho_\mathrm{b}$) and bulk viscosities ($\mu_\mathrm{a}, \mu_\mathrm{b}$). Fluid $a$ is referred to as foreground fluid and fluid $b$ as background fluid. In the present study, the foreground fluid typically refers to air and the background fluid to water. Both fluids are assumed to share the kinematic field along the route of the VoF-approach suggested by \cite{hirt1981volume}. The spatial distribution of the fluids is  described by an Eulerian concentration field, where $c=c_\mathrm{a}=V_\mathrm{a}/V$ denotes the volume concentration of the foreground fluid, and the  volume fraction occupied by the background fluid refers to $c_\mathrm{b}=V_\mathrm{b}/V=(V-V_\mathrm{a})/V=(1-c)$.

\subsubsection{Concentration Transport}
\label{sec:conctrans}
The material properties of immiscible and inert fluids are invariable. The (foreground) fluid concentration of a VoF model, therefore, follows from a simple Lagrangian transport equation, i.e. $\mathrm{d} c_\mathrm{a}/ \mathrm{d} t  \, (=- \mathrm{d} c_\mathrm{b}/ \mathrm{d} t) = \mathrm{d} c/\mathrm{d} t =0$, which is translated into an Eulerian formulation prior to its discretization. More elaborate diffuse interface methods exist, which are frequently labeled Cahn-Hilliard (CH) models, cf. \cite{cahn1958free, lowengrub1998quasi, jacqmin1999calculation, abels2012thermodynamically}. CH models replace the sharp interface with a thin layer where the fluids exchange mass fluxes. They are distinguished by mass or volume conservative strategies and essentially augment the Lagrangian concentration transport equation by a velocity-divergence term and a non-linear, diffusive right-hand side of order four, which is zero outside the interface region, cf. \cite{ding2007diffuse} and \cite{kuhl2021cahn} 
\begin{align}
\frac{\mathrm{d} c}{\mathrm{d} t} = 
   \frac{\partial}{\partial x_\mathrm{k}}
  \left[ M \frac{\partial \psi}{\partial x_\mathrm{k}} 
  \right] - c \frac{\partial v_\mathrm{k}}{\partial x_\mathrm{k}} 
  \quad \to \quad
     \frac{\partial c}{\partial t} + \frac{\partial \, v_\mathrm{k} c}{\partial x_\mathrm{k}} =
      \frac{\partial}{\partial x_k}
  \left[ M \frac{\partial \psi}{\partial x_k} 
  \right] 
  \; . 
  \label{equ:CHceq}
\end{align}
Here $M(c)$ refers to a mobility parameter of dimension [m$^4$/(N s)] and $\psi(c, \Delta c)$ denotes to a chemical potential of dimension [Pa]. Following \cite{kuhl2021cahn}, the present study employs a mass conservative strategy together with an appropriate choice of $M$ and a frequently used ''double-well potential'' which yields
\begin{align}
  \psi = 2 C_\mathrm{1} \left[ (2c^3 -3c^2+c) - 0.5 \left(\frac{C_\mathrm{2}}{C_\mathrm{1}}\right) \frac{\partial^2 c}{\partial x_k^2} \right]\;   \quad \to \quad 
  \frac{\partial \psi}{\partial x_k}  = 2 C_\mathrm{1} \left[ (6c^2 -6c+1) \frac{\partial c}{\partial x_k}  - 0.5 \left(\frac{C_\mathrm{2}}{C_\mathrm{1}}\right) \frac{\partial^3 c}{\partial x_k^3} \right]\;   .
  \label{equ:CHceqB}
\end{align}
The ratio $C_\mathrm{2} [N]/C_\mathrm{1} [Pa] \sim \gamma_\mathrm{c}^2$ scales with the square of the interface thickness, and $C_\mathrm{1} \cdot M [m^2/s] \sim \nu_c$ describes a nonlinear apparent viscosity $\nu_\mathrm{c} = 2 C_\mathrm{1} \cdot M (6c^2 -6c+1)$. Evaluating the  last term of $\nabla_k \psi$ requires sufficient grid resolution, in other words, the term can be neglected when the interface is under-resolved, which is the case in marine engineering simulations. As illustrated in Fig.  \ref{fig:double_well_potential}, $\nu_c$ vanishes at $c=(0.5 \pm \sqrt{3}/6)$  and is negative over approximately 58\% of the inner transition regime, where it supports the phase separation process. 

Though the non-zero RHS of (\ref{equ:CHceq}) appears to increase the complexity, it is beneficial for various reasons, cf. \cite{kuhl2021cahn}: It naturally includes surface tension effects, supports the use of stability-preserving, upwind-biased convective approximations, and facilitates consistent yet robust primal/adjoint formulations. The latter is particularly relevant for the present study.

\subsubsection{Equation of State}
An equation of state (EoS) $m(c)$ extracts the local flow properties from the concentration field and the bulk properties, viz. 
\begin{align}
\rho = m^\mathrm{\rho}  \rho^\mathrm{\Delta} + \rho_\mathrm{b} 
\qquad \qquad \mathrm{and} \qquad \qquad
\mu = m^\mathrm{\mu}  \mu^\mathrm{\Delta} + \mu_\mathrm{b} \, , \label{equ:mater_prope}
\end{align}
where $\rho^\mathrm{\Delta} = \rho_\mathrm{a} - \rho_\mathrm{b}$, $\mu^\mathrm{\Delta} = \mu_\mathrm{a} - \mu_\mathrm{b}$ mark the respective bulk property differences. Though this is not necessary, the paper assigns $m^\mathrm{\mu} = m^\mathrm{\rho}$.
Provisions on the EoS considered in this study aim to exclude non-physical, unbounded density states by means of $m \in \left[0,1\right]$ and to recover the single-phase limit states, i.e. $m(c=1[0]) = 1[0]$, cf.  \cite{kuhl2021cahn, kuhl2021phd}.
The simplest conceivable EoS $m^{(1)}$ corresponds to a bounded linear interpolation between the limit states. A more advanced nonlinear alternative $m^{(2)}$ follows the rule of a hyperbolic tangent and employs a user-specified non-dimensional transition parameter $\gamma^\mathrm{m}$ 
\begin{align}
m^\mathrm{(1)} = \begin{cases}
0 &\text{if} \ c < 0 \\
1 &\text{if} \ c > 1 \\
c &\text{otherwise}
\end{cases}
\qquad \qquad \mathrm{and} \qquad \qquad
m^\mathrm{(2)} = \frac{1}{2} \left[ \mathrm{tanh} \left( \frac{2 c - 1}{\gamma^\mathrm{m}}\right) +1 \right] \, . \label{equ:eos_general}
\end{align}
Since the hyperbolic EoS complies with the limit states only asymptotically, an upper bound for the transition parameter is estimated by $\gamma^\mathrm{m} \le 0.3$  to limit the error w.r.t. the limit states below 0.1\%. Typical values for the transition parameter refer to $0.25 \le \gamma^\mathrm{m} \le 0.35$. 

In combination with a CH-formulation, the hyperbolic EoS offers a decisive advantage for constructing a consistent primal/adjoint two-phase flow model dedicated to shape optimization, which closely resembles the traditional VoF framework. The benefit applies to academic studies featuring grid-resolved interface physics and, even more importantly, engineering simulations with under-resolve interface physics. 
Introducing the EoS (\ref{equ:mater_prope}) into the mass conservative continuity equation yields an expression for the divergence of the velocity field that is essentially governed by (\ref{equ:eos_general}) even for a diffuse interface scheme.
\begin{align}
    \frac{\partial \rho}{\partial t} + \frac{\partial \, v_\mathrm{k} \rho}{\partial x_\mathrm{k}} = 0
    %\sigma_\mathrm{a} + \sigma_\mathrm{b}
    \qquad \qquad \to \qquad \qquad
    \frac{\partial v_\mathrm{k}}{\partial x_\mathrm{k}} =
     \frac{\sigma_\mathrm{a}}{\rho_\mathrm{a}}+ \frac{\sigma_\mathrm{b}}{\rho_\mathrm{b}}
      = f^\mathrm{\rho} \frac{\mathrm{d c}}{d t} 
    \qquad \mathrm{with} \qquad
    f^\mathrm{\rho} = \frac{- \rho^\mathrm{\Delta}}{\rho} \frac{\partial \, m}{\partial \, c} \, . \label{equ:primal_mass_conservation_final}
\end{align}
Here $\sigma_\mathrm{a} = -\sigma_\mathrm{b}$ represent the mass transfer rates into phases $a$ and $b$. Mass conservative  CH formulations yield non-solenoidal velocity fields unless $f^\rho$ vanishes, cf. \cite{kuhl2021cahn}. This in turn suggests to employ the hyperbolic EoS which can compress the non-solenoidal regime to a small layer controlled by $\gamma^\mathrm{m}$.

\subsection{Primal Governing Equations}
The governing fluid dynamic equations refer to the momentum and continuity equation for the mixture as well as a transport equation for the volume concentration of the foreground phase, that need to be solved for the pressure $p$, the velocity $v_\mathrm{i}$, and the concentration $c$,
viz.
\begin{alignat}{3}
&\mathrm{R}^\mathrm{ p}=&&  \frac{\partial v_\mathrm{k}}{\partial x_\mathrm{k}}  - \frac{f^\mathrm{\rho}}{1 + f^\mathrm{\rho} \, c} \,  \frac{\partial}{\partial \, x_\mathrm{k}} \bigg[ M \frac{\partial \, \psi}{\partial \, x_\mathrm{k}}\bigg] && = 0 \label{equ:primal_rans_masse_unlucky} \\
&\mathrm{R}^\mathrm{ c}=&&  \frac{\mathrm{d} \, c}{\mathrm{d} \, t} - \frac{1}{1 + f^\mathrm{\rho} \, c} \,  \frac{\partial}{\partial \, x_\mathrm{k}} \bigg[ M \frac{\partial \, \psi}{\partial \, x_\mathrm{k}}\bigg] &&= 0 \label{equ:primal_rans_concee_unlucky} \\
&\mathrm{R}_\mathrm{i}^\mathrm{ v_\mathrm{i}}=&& \rho \frac{\mathrm{d} \, v_\mathrm{i}}{\mathrm{d} \, t} + \frac{\partial }{\partial \, x_\mathrm{k}} \bigg[p^\mathrm{ eff}  \, \delta_\mathrm{ik} - 2 \, \mu^\mathrm{ eff} \, S_\mathrm{ik} \bigg] - \rho \, g_\mathrm{i} + \frac{2}{3} \frac{\partial}{\partial \, x_\mathrm{i}} \bigg[ \mu \, \frac{f^\mathrm{\rho}}{1 + f^\mathrm{\rho} \, c} \frac{\partial}{\partial \, x_\mathrm{k}} \bigg[ M \frac{\partial \, \psi}{\partial \, x_\mathrm{k}}\bigg] \bigg] &&=0 \, . \label{equ:primal_rans_mome_unlucky} 
\end{alignat}
The unit coordinates and the strain rate tensor are denoted by $\delta_\mathrm{ik}$ and $S_\mathrm{ik}$.
The framework supports laminar and Reynolds-averaged (modeled) turbulent flows (RANS). In the latter case, $v_\mathrm{i}$ and $p^\mathrm{eff}$ correspond to Reynolds-averaged properties and $p^\mathrm{eff}$ is additionally augmented by a turbulent kinetic energy ($k$) term, i.e. $2 \rho k/3$. Along with the Boussinesq hypothesis, the dynamic viscosity $\mu^\mathrm{eff} = \mu + \mu_\mathrm{t}$ of turbulent flows consists of a molecular and a turbulent contribution ($\mu_\mathrm{t}$), and the system is closed using a two-equation turbulence model to determine $\mu_\mathrm{t}$ and $k$. Details of the turbulence modeling practice are omitted to safe space and can be found in textbooks, e.g., \cite{wilcox1998turbulence}.

\begin{figure}[!ht]
\centering
\iftoggle{tikzExternal}{
\subfigure[]{\input{./tikz/double_well_potential.tikz}}}{
\subfigure[]{\includegraphics{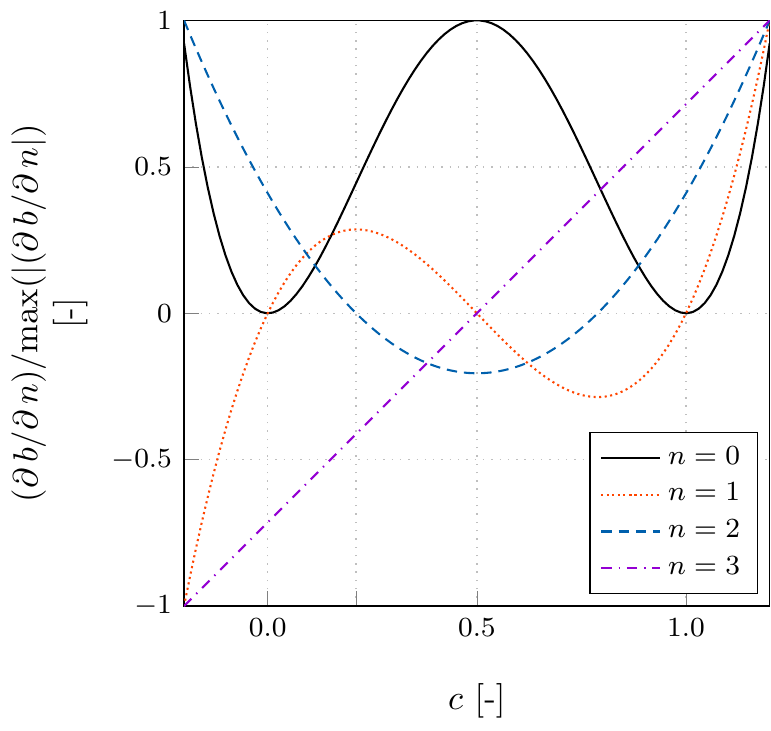}}
}
\iftoggle{tikzExternal}{
\subfigure[]{\input{./tikz/apparent_viscosity.tikz}}}{
\subfigure[]{\includegraphics{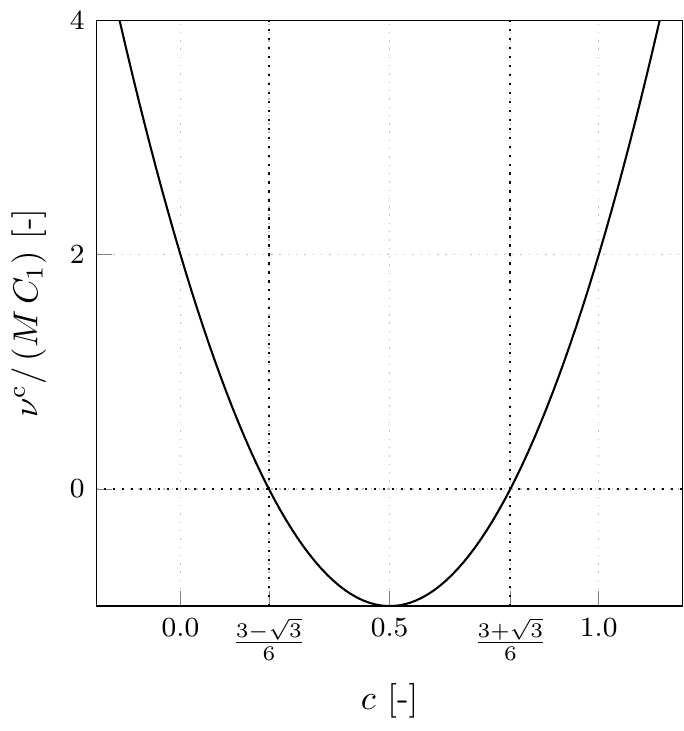}}
}
\caption{Double well potential: (a) Normalized fourth-order polynomial $b$ as well as its first three (normalized) derivatives and (b) the evolution of the normalized apparent viscosity with indicated roots.}
\label{fig:double_well_potential}
\end{figure}

Contributions arising from the two-phase model are noted in the respective final positions of Eqns. (\ref{equ:primal_rans_masse_unlucky})-(\ref{equ:primal_rans_mome_unlucky}). The PDE system agrees with the classical VoF framework for a vanishing mobility $M \to 0$. A divergence-free velocity field (\ref{equ:primal_rans_masse_unlucky}) is often highly appreciated and also reduces the differentiation efforts during the subsequent derivation  of a continuous adjoint formulation. Using the nonlinear material model  $m^\mathrm{(2)}$ in (\ref{equ:eos_general}), $f^\mathrm{\rho}$ approximately vanishes due to $\partial m / \partial c \to 0$ for sufficiently small values of $\gamma^\mathrm{m}$. Whilst this yields the neglect of  net diffusion fluxes in (\ref{equ:primal_rans_masse_unlucky}) and surface tension effects in (\ref{equ:primal_rans_mome_unlucky}), it leaves a diffusive term within the concentration equation (\ref{equ:primal_rans_concee_unlucky}). The latter arises from the first part of the chemical potential $\psi$, cf. Eqn.  (\ref{equ:CHceqB}), and yields a consistent -- therefore robust -- adjoint formulation. Such under-resolved CH-VoF methods consistently employ $f^\mathrm{\rho} \to 0 $ to simplify the primal PDE system, and serve as the basis of our adjoint two-phase flow derivation, viz.
\begin{alignat}{3}
&\mathrm{R}^\mathrm{ p} &&= \frac{\partial v_\mathrm{k}}{\partial x_\mathrm{k}} &&= 0 \label{equ:primal_rans_mass} \\
%%%
&\mathrm{R}^\mathrm{ c} &&= \frac{\partial \, c}{\partial \, t} + \frac{\partial \, v_\mathrm{k} \, c}{\partial x_\mathrm{k}} - \frac{\partial}{\partial x_\mathrm{k}} \left[ \nu_\mathrm{c} \frac{\partial c}{\partial x_\mathrm{k}} \right] &&= 0 \label{equ:primal_rans_concentration} \\
%%%
&\mathrm{R}_\mathrm{i}^\mathrm{ v_\mathrm{i}} &&= \rho \bigg[ \frac{\partial \, v_\mathrm{i}}{\partial \, t} + v_\mathrm{k} \frac{\partial \, v_\mathrm{i}}{\partial \, x_\mathrm{k}} \bigg] + \frac{\partial }{\partial \, x_\mathrm{k}} \bigg[p^\mathrm{ eff}  \, \delta_\mathrm{ij} - 2 \, \mu^\mathrm{ eff} \, S_\mathrm{ij} \bigg] - \rho \, g_\mathrm{i} &&=0 \label{equ:primal_rans_momentum} \, .
\end{alignat}
The nonlinear apparent viscosity reads $\nu_\mathrm{ c} = M \partial^2 b/\partial c^2 = 2 C_\mathrm{1} \, M (6 \, c^2 - 6 \, c + 1)$, cf. Sec.  \ref{sec:conctrans}. As indicated by Fig. \ref{fig:double_well_potential} (left), it follows from a double-well potential $b = (c-1)^2 c^2$ to be minimized in a phase separation process. Depending on the concentration value $c$, the last term in (\ref{equ:primal_rans_concentration}) acts locally diffusive ($\nu_\mathrm{ c} \ge 0$) or compressive ($\nu_\mathrm{ c} <0$), cf. Fig. \ref{fig:double_well_potential} right. This underlines the compressive character of the CH-VoF approach. The primal two-phase flow model is closed by assigning the product $C_\mathrm{1} M$ to a spatially constant value that is guided by the numerical diffusion of the primal convective concentration transport as suggested by \cite{kuhl2021cahn}. 

Boundary conditions to close the PDE system (\ref{equ:primal_rans_mass})-(\ref{equ:primal_rans_momentum}) are listed in Tab. \ref{tab:primal_boundary_conditions}. Constant values of the pressure gradient typically follow from a piezometric effective pressure definition, viz. $p^\mathrm{ eff} \to p + \rho \, x_\mathrm{k} \, g_\mathrm{k}$ and thus $\partial p^\mathrm{ eff} / \partial x_\mathrm{i} \to \partial p / \partial x_\mathrm{i} + \rho \, g_\mathrm{i}$. In all cases, inlet, symmetry and slip-wall boundaries are perpendicular to the gravity vector and no hydrostatic boundary contributions occur.

\begin{table}[!ht]
\begin{center}
\begin{tabular}{|c||c|c|c|}
\hline
boundary type & $v_\mathrm{i}$  & $p$ & $c$ \\
\hline
\hline
inlet & $v_\mathrm{i} = v_\mathrm{i}^\mathrm{ in}$ & $\frac{\partial p}{\partial n} = 0$ & $c = c^\mathrm{in}$  \\
\hline
outlet & $\frac{\partial v_\mathrm{i}}{\partial n} = 0$ & $p = \rho \, g_\mathrm{k} \, x_\mathrm{k}$ & $\frac{\partial c}{\partial n} = 0$  \\
\hline
symmetry & $v_\mathrm{i} \, n_\mathrm{i} = 0$, $\frac{\partial \, v_\mathrm{i}}{\partial \, n} t_\mathrm{i} $ & $\frac{\partial p}{\partial n} = 0$ & $\frac{\partial c}{\partial n} = 0$ \\
\hline
wall (slip) & $v_\mathrm{i} \, n_\mathrm{i} = 0$, $\frac{\partial \, v_\mathrm{i}}{\partial \, n} t_\mathrm{i} $ & $\frac{\partial p}{\partial n} = 0$ & $\frac{\partial c}{\partial n} = 0$ \\
\hline
wall (no-slip) & $v_\mathrm{i} = v_\mathrm{i}^\mathrm{ w}$
& $\frac{\partial p}{\partial n}  = \rho \, g_\mathrm{k} \, n_\mathrm{k}$ & $\frac{\partial c}{\partial n} = 0$ \\
\hline
\end{tabular}
\end{center}
\caption{Boundary conditions for the primal equations, where $t_\mathrm{i}$ [$n_\mathrm{i}$] refer to the local boundary tangential [normal] vector.}
\label{tab:primal_boundary_conditions}
\end{table}

\subsection{Adjoint Governing Equations} 
The adjoint PDE system depends on an underlying integral objective functional, viz.
\begin{align}
J = \int_{\Omega_\mathrm{O}} j_\Omega \, \mathrm{d} \Omega  + \int_{\Gamma_\mathrm{O}} j_\Gamma \, \mathrm{d} \Gamma \; ,  \label{equ:general_objective}
\end{align}
that either acts in parts of the domain ($\Omega_\mathrm{O} \subseteq \Omega $) or along boundary segments ($\Gamma_\mathrm{O} \subseteq \Gamma$ ). Both integrands in (\ref{equ:general_objective}) can depend on the field quantities of the primal system  (\ref{equ:primal_rans_mass})-(\ref{equ:primal_rans_momentum}). The objectives used in this paper read
\begin{align}
j_\Omega =   \frac{1}{2} \left[ c - c_\mathrm{t} \right]^2 
\qquad \qquad \mathrm{and} \qquad \qquad 
j_\Gamma =  \left[ p \delta_{\mathrm{ij}} - 2 \mu S_{\mathrm{ij}} \right] n_\mathrm{j} r_\mathrm{i} \, . \label{equ:special_objective}
\end{align}
The volume objective minimizes the deviation from a target concentration value $c_\mathrm{t}$, e.g. aiming at calm water elevation. The surface objective addresses the fluid force projected in a spatial direction $r_\mathrm{i}$, e.g. the fluid flow induced drag.
The cost functional (\ref{equ:general_objective}) is augmented by the  primal PDE system (\ref{equ:primal_rans_mass})-(\ref{equ:primal_rans_momentum}) which yields the following Lagrangian
\begin{align}
L = J + \int \int \left[  \hat{p} \, \mathrm{R}^\mathrm{ p} + \hat{c} \,\mathrm{R}^\mathrm{ c} + \hat{v}_\mathrm{i} \,\mathrm{R}^\mathrm{ v_i} \right] \mathrm{d}\Omega \, \mathrm{d} t  \label{equ:simplify_derive_lagragian_rans} \, .
\end{align}
In (\ref{equ:simplify_derive_lagragian_rans}) $\hat{p}$, $\hat{c}$ and $\hat{v}_\mathrm{i}$ refer to adjoint pressure, adjoint concentration and adjoint velocity components, respectively. The units of adjoint pressure and adjoint concentration are equal $\left[ \hat{p} \right] = \left[ \hat{c} \right] = [J] \, 1  / \mathrm{m}^3$. The unit of the adjoint velocity read $\left[ \hat{v}_i \right] = [J] \, 1  / \mathrm{(N \, s)}$. 
Demanding first-order optimality conditions yields $\delta_{\hat \phi, \phi, u}L = 0$ (\cite{kuhl2019decoupling, kuhl2021continuous}), where $\hat \phi\in[\hat p, \hat c, \hat v_i]$ and $\phi\in[p, c, v_i]$ denote to the adjoint and primal variables and $u$ 
refers to the control, i.e. the normal displacement of discrete surface-element's centroids for a shape optimization purposes. The optimality conditions yield three sets of constraints.
The first part ($\delta_{\hat \phi}L = 0$) reproduces the PDE system 
(\ref{equ:primal_rans_mass})-(\ref{equ:primal_rans_momentum}).
 The second part
($\delta_{\phi}L = 0$)  
 yields the adjoint field equations, viz.
\begin{alignat}{3}
&\mathrm{R}^\mathrm{ \hat{p}} &&= -\frac{\partial \hat{v}_\mathrm{k}}{\partial x_\mathrm{k}} + \frac{\partial \, j^\mathrm{ \Omega}}{\partial \, p} &&= 0  \label{equ:adjoint_rans_mass} \\
%%%
&\mathrm{R}^\mathrm{ \hat{c}} &&=  - \frac{\partial \, \hat{c}}{\partial \, t} - v_\mathrm{k} \frac{\partial \, \hat{c} }{\partial \, x_\mathrm{k}} - \rho^\mathrm{ \Delta} \bigg[ \hat{v}_\mathrm{i} \, g_\mathrm{i} - \hat{v}_\mathrm{i} \, v_\mathrm{k} \frac{\partial \, v_\mathrm{i}}{\partial \, x_\mathrm{k}} - \frac{\partial \, \mu^\mathrm{ t}}{\partial \, \rho} S_\mathrm{ik} \frac{\partial \, \hat{v}_\mathrm{i}}{\partial \, x_\mathrm{k}} \bigg] \frac{\partial m}{\partial c} && \nonumber \\
& && \qquad \qquad \qquad \quad \ \ + \mu^\mathrm{ \Delta} \bigg[  2 \, S_\mathrm{ik} \frac{\partial \, \hat{v}_\mathrm{i}}{\partial \, x_\mathrm{k}} \bigg] \frac{\partial m}{\partial c} + \frac{\partial \, j^\mathrm{ \Omega}}{\partial \, c} - \frac{\partial}{\partial \, x_\mathrm{k}} \bigg[ \nu_\mathrm{\hat{c}} \frac{\partial \, \hat{c}}{\partial \, x_\mathrm{k}} \bigg]&&= 0 \label{equ:adjoint_rans_concentration} \\
%%%
&\mathrm{R}_\mathrm{i}^\mathrm{ \hat{v}_\mathrm{i}} &&= -\rho \bigg[ \frac{\partial \, \hat{v}_\mathrm{i}}{\partial \, t} + v_\mathrm{k} \frac{\partial \, \hat{v}_\mathrm{i}}{\partial \, x_\mathrm{k}} \bigg] + \frac{\partial }{\partial \, x_\mathrm{k}} \bigg[\hat{p} \, \delta_\mathrm{ik} - 2 \, \left(\mu + \beta \mu^\mathrm{t} \right) \, \hat{S}_\mathrm{ik} \bigg] + \hat{c} \frac{\partial \, c}{\partial \, x_\mathrm{i}} + \rho \, \hat{v}_\mathrm{k}  \frac{\partial \, v_\mathrm{k}}{\partial \, x_\mathrm{i}} + \frac{\partial \, j^\mathrm{ \Omega}}{\partial \, v_\mathrm{i}} &&= 0 \label{equ:adjoint_rans_momentum} \, ,
\end{alignat}
that are supplemented by boundary integrals
\begin{align}
\delta_\mathrm{p} L 
%\cdot \delta p 
&= 
\int_\mathrm{\Gamma_\mathrm{O}} \frac{\partial \, j_\mathrm{ \Gamma}}{\partial \, p} \delta p\, \mathrm{d} \Gamma 
+ \int_\mathrm{\Omega_\mathrm{O}} \frac{\partial \, j_\mathrm{ \Omega}}{\partial \, p} \delta p \, \mathrm{d} \Omega \nonumber \\
&\qquad + \int \int \delta p \, \hat{v}_\mathrm{i} n_\mathrm{i} \, \mathrm{d} \Gamma \mathrm{d} t \quad \overset{!}{=} 0 \quad  \forall \, \delta p \label{equ:adjoint_rans_mass_bc} \\
\delta_\mathrm{c} L 
% cdot \delta c 
&= 
\int_\mathrm{\Gamma_\mathrm{O}} \frac{\partial \, j_\mathrm{\Gamma}}{\partial \, c} \delta c \, \mathrm{d} \Gamma 
+ \int_\mathrm{\Omega_\mathrm{O}} \frac{\partial \, j_\mathrm{\Omega}}{\partial \, c} \delta c  \, \mathrm{d} \Omega \nonumber \\
&\qquad+ \int \int \delta c \bigg[ \hat{c} \, v_\mathrm{k} \, n_\mathrm{k} + \nu^\mathrm{ c} \frac{\partial \, \hat{c}}{\partial \, n} + \hat{v}_\mathrm{i} \bigg[ \frac{2}{3}  k \, n_\mathrm{i} - 2 \, \frac{\partial \, \mu^\mathrm{ t}}{\partial \, \rho} S_\mathrm{ik} \, n_\mathrm{k} \bigg] \rho^\mathrm{ \Delta} \frac{\partial m}{\partial c} 
%m^\mathrm{\rho \prime}
%\nonumber \\
%& \qquad \qquad \qquad 
- \hat{v}_\mathrm{i} \, 2 \, S_\mathrm{ik} \, n_\mathrm{k} \, \mu^\mathrm{\Delta} 
%m^\mathrm{\mu \prime} 
\frac{\partial m}{\partial c} \bigg] \nonumber \\
& \qquad \qquad  - \frac{\partial \, \delta c}{\partial \,n} \bigg[ \hat{c} \, \nu^\mathrm{c} \bigg] \mathrm{d} \Gamma \, \mathrm{d} t \quad \overset{!}{=} 0 \quad  \forall \, \delta c \label{equ:adjoint_rans_concentration_bc} \\
\delta_\mathrm{v_\mathrm{i}} L 
%\cdot \delta v_\mathrm{i} 
&=
\int_\mathrm{\Gamma^\mathrm{O}}  \frac{\partial \, j^\mathrm{ \Gamma}}{\partial \, v_\mathrm{i}} \delta v_\mathrm{i} \, \mathrm{d} \Gamma 
+ \int_\mathrm{\Omega^\mathrm{O}}  \frac{\partial \, j^\mathrm{ \Omega}}{\partial \, v_\mathrm{i}} \delta v_\mathrm{i}  \, \mathrm{d} \Omega \nonumber \\
&\qquad+ \int \int \delta v_\mathrm{i} \bigg[ v_\mathrm{k} \, \rho \, \hat{v}_\mathrm{i} \, n_\mathrm{k} + 2 \mu^\mathrm{ eff} \hat{S}_\mathrm{ik} \, n_\mathrm{k} -\hat{p} \, n_\mathrm{i} \bigg] - \delta S_\mathrm{ik} \bigg[ 2 \mu^\mathrm{ eff} \, \hat{v}_\mathrm{i} \, n_\mathrm{k} \bigg] \mathrm{d} \Gamma \mathrm{d} t \quad \overset{!}{=} 0 \quad  \forall \, \delta v_\mathrm{i} \label{equ:adjoint_rans_momentum_bc} \, .
\end{align}
The third part $(\delta_\mathrm{u}L=0)$ provides the sensitivity that guides the shape update. 

All derivatives of material properties enter the adjoint concentration equation with the individual differences of the bulk properties $\rho^\mathrm{\Delta}$ and $\mu^\mathrm{\Delta}$, which are multiplied by the derivative $\partial m/\partial c$ of the EoS. The linear EoS $m^{(1)}$ offers a constant unit derivative $\partial m/\partial c =1$, and its non-linear alternative reveals an intensified local contribution along the interfacial region that weakens noticeably towards the bulk phases. For a vanishing thickness parameter $\gamma^m$, the hyperbolic tangent in (\ref{equ:eos_general}) turns into a Heaviside function, and the adjoint system experiences an abrupt (Dirac) impulse along the interface based on the EoS related source terms in (\ref{equ:adjoint_rans_concentration}). However, the integral impact does not change and exactly matches that of a linear approach. Such thought experiments reveal the vulnerability of the discrete framework: The adjoint system conceptually pushes the phase transition below the grid resolution in practical applications. Hence, in adjoint mode we consequently apply a linear EoS, i.e. the use of $\partial m/\partial c =1$ in (\ref{equ:adjoint_rans_mass})-(\ref{equ:adjoint_rans_momentum_bc}), in order not to compromise the numerical robustness.
Comparing the primal and the adjoint PDE systems, a few additional advection and cross-coupling terms occur in the adjoint PDE system. Above all, the adjoint concentration equation contains significantly more terms that scale with the two fluids' bulk density or bulk viscosity difference. The last term of (\ref{equ:adjoint_rans_concentration}) is of particular importance since this additional diffusivity bridges the gap between an adjoint sharp vs. an under-resolved diffusive interface formulation. We would like to point out that the differentiation of the non-linear material model (\ref{equ:eos_general}) and the differentiation of the apparent viscosity $\nu_c$ were deliberately suppressed. They would yield additional contributions to the adjoint concentrations equation, which are proportional to $\sim (2c-1) \partial c/\partial x_k$ and are thus confined to two small regions along with the phase transition regime which are separated in the vicinity of the interface, i.e., at $c=0.5$. Moreover, the treatment of the adjoint apparent viscosity $\nu_\mathrm{\hat{c}}$  is also simplified and assigned to a spatially constant, positive and thus stability-promoting value that follows from the bulk phase, i.e. $\nu_\mathrm{\hat{c}} = 2 M \, C_1$, cf. Fig. \ref{fig:double_well_potential} right.
Boundary conditions to close the adjoint PDE system (\ref{equ:adjoint_rans_mass})-(\ref{equ:adjoint_rans_momentum}) aim at neutralizing the boundary integrals (\ref{equ:adjoint_rans_mass_bc})-(\ref{equ:adjoint_rans_momentum_bc}) and are listed in Tab. \ref{tab:adjoint_boundary_conditions}.
\begin{table}[!ht]
\begin{center}
\begin{tabular}{|c||c|c|c|}
\hline
boundary type & $\hat{v}_\mathrm{i}$  & $\hat{p}$ & $\hat{c}$ \\
\hline
\hline
inlet & $\hat{v}_\mathrm{i} = 0$ & $\frac{\partial \, \hat{p}}{\partial \, n} = 0$ & $\hat{c} = 0$  \\
\hline
outlet & $\frac{\partial \, \hat{v}_\mathrm{i}}{\partial \, n} = 0$ & $\hat{p} = [ v_\mathrm{k} \, \rho \, \hat{v}_\mathrm{i} + \mu^\mathrm{eff} \hat{S}_\mathrm{ik} \, n_\mathrm{k} ] \, n_\mathrm{i}$ & $\hat{c} = 0$  \\
\hline
symmetry & $\hat{v}_\mathrm{i} \, n_\mathrm{i} = 0$, $\frac{\partial \, \hat{v}_\mathrm{i}}{\partial \, n} t_\mathrm{i} = 0$ & $\frac{\partial \, \hat{p}}{\partial \, n} = 0$ & $\frac{\partial \, \hat{c}}{\partial \, n} = 0$ \\
\hline
wall (slip) & $\hat{v}_\mathrm{i} \, n_\mathrm{i} = 0$, $\frac{\partial \, \hat{v}_\mathrm{i}}{\partial \, n} t_\mathrm{i} = 0$ & $\frac{\partial \, \hat{p}}{\partial \, n} = 0$ & $\frac{\partial \, \hat{c}}{\partial \, n} = 0$ \\
\hline
\hline
wall (no-slip, $\Gamma \not\subset \Gamma^\mathrm{O}$) & $\hat{v}_\mathrm{i} = 0$ & $\frac{\partial \, \hat{p}}{\partial \, n} = 0$ & $\frac{\partial \, \hat{c}}{\partial \, n} = 0$ \\
\hline
wall (no-slip, $\Gamma \subset \Gamma^\mathrm{O}$) & $\hat{v}_\mathrm{i} = -r_\mathrm{i}$ & $\frac{\partial \, \hat{p}}{\partial \, n} = 0$ & $\frac{\partial \, \hat{c}}{\partial n} = 0$ \\
\hline
\end{tabular}
\end{center}
\caption{Boundary conditions for the adjoint equations, where $t_\mathrm{i}$ [$n_\mathrm{i}$] refer to the local boundary tangential [normal] vector.}
\label{tab:adjoint_boundary_conditions}
\end{table}

\subsubsection*{Adjoint Turbulence Treatment}
The adjoint turbulence treatment follows from the suggestion published in \cite{kuhl2021adjoint_2}, which recommends that the adjoint equations employ  twice the primal turbulent viscosity $\beta = 2$ [$\beta = 1$] in a ''weakly consistent'' [frozen] case. Mind that $\beta = 2$ is consistent in the viscous sub-layer and the logarithmic region. At the same time, it can only be hypothesized that the consistency improves compared to the frozen turbulence approach for other applications. However, shape optimization for resistance problems is by definition interested in the primal / adjoint near-wall flow. Hence a consistent adjoint formulation should be particularly relevant in this region. Moreover, the robustness of the adjoint numerical procedure benefits from an augmented viscosity.

\subsubsection*{Adjoint Sensitivity}
If all respective optimality conditions are satisfied, sensitivity information are obtained from the final optimality condition in terms of a derivative of the Lagrangian in the direction of the control, i.e. $\delta_u L$. The latter gives rise to the desired shape sensitivity derivative $s$, which follows from $(\partial \, v_\mathrm{i} / \partial \, n) \, n_\mathrm{i} = 0$ and reads
\begin{align}
\delta_\mathrm{u} L = - \int \int_\mathrm{\Gamma^{ D}} (\mu + \beta \, \mu^\mathrm{ t}) \frac{\partial \, v_\mathrm{i}}{\partial \, n} \left[ - \frac{\partial \, \hat{v}_\mathrm{i}}{\partial \, n} \right] \mathrm{d} \Gamma \mathrm{d} t 
\qquad \to \qquad s = -(\mu + \beta \, \mu^\mathrm{ t}) \frac{\partial \, v_\mathrm{i}}{\partial \, n}
\left[ \frac{\partial \, \hat{v}_\mathrm{i}}{\partial \, n}
\right]
\, . \label{equ:shape_derivative}
\end{align}

\subsubsection*{Interpretation of Primal vs. Dual Time Horizon} 
Being primarily concerned with the drag reduction of ships cruising in calm water, the present research focuses on steady-state problems. Therefore, the primal and adjoint solutions are advanced in pseudo-time and converged to a steady-state. To this end, all adjoint time steps are solely linearized around the final (steady-state) primal flow solution. A pseudo-transient procedure also influences the identification of the floatation position outlined in Sec. \ref{sec:gridupdate} which does not need to consider inertia effects of the rigid body mechanics. 

%% file: tex/grid_modifications.tex
Both the adjustment of the floatation and the shape modification suggested by the optimizer employ the same template to update the numerical grid along the lines of a mesh morphing procedure to facilitate a restart from the previous design. In both cases, the grid update procedure is driven by the spatial change of the discretized vessel geometry. 

\subsection{Modeling of Floatation}
\label{subsec:floatation}
The paper considers an adjustment of the trim and sinkage during the integration to steady-state, using a rigid body motion model (\cite{luo2017computation})  that is restricted to two degrees of freedom herein. Due to the pseudo transient approach, inertia aspects are irrelevant for the final floatation, and a hydrostatic approach to adjust the floatation can be pursued, cf. \cite{yang2002calculation}. The floating position is initialized in its hydrostatic rest position in this simplified approach. This rest position is associated with an initial displacement  $V^\mathrm{ ini}$ and supplemented by centers of gravity $ x_\mathrm{k}^\mathrm{ g}$ and rotation $ x_\mathrm{k}^\mathrm{ r}$. The initial displacement corresponds to the gravity neutralizing buoyancy force and is associated with the vessel's carrying capacity. The latter is conserved during the optimization in the present study, cf. Sec. \ref{subsec:geometrical_constraints}. Once the flow develops, the related forces deviate from the initial or the previous iteration -- possibly due to a modified shape -- and the floatation is corrected. To this end, the surface grid of the hull, which refers to the interior boundary of the discrete domain, is rigidly displaced and rotated. In contrast, all exterior boundaries remain unmoved, as outlined in Sec. \ref{sec:meshdeform}. 

For the sake of simplicity, we assume the gravitation and heave force to act in the negative $x_\mathrm{3}$ (vertical) direction. Consequently, the trim moment is associated with the pitch-down positive $x_\mathrm{2}$ (span) rotation, perpendicular to the cruise direction ($x_\mathrm{1}$) and the gravity vector $g_\mathrm{k}$. An estimation of the required trim (pitch rotation) and sinkage (vertical motion) correction w.r.t. the initial hydrostatic floatation follows from the actual trim moment $M^\mathrm{ T}$ and the net heave force $F^\mathrm{ H}$, viz. 
\begin{align}
\Delta S_\mathrm{3}[m] = - \frac{F^\mathrm{ H}}{\rho^\mathrm{ b} \, |g_\mathrm{k}| \, A^\mathrm{ w} } \; %\underline e_\mathrm{3}
\qquad \mathrm{and} \qquad
\Delta T_\mathrm{2}[^\circ] = \frac{M^\mathrm{ T}}{\rho^\mathrm{ b} \, |g_\mathrm{k}| \, I^\mathrm{ w} } \; 
%\underline e_\mathrm{2}
\; . 
\label{equ:adaptive_trim_sinkge_error}
\end{align}
Here $A^\mathrm{ w}$ and $I^\mathrm{ w}$ represent the water-plane area and its moment of inertia around the rotating axis in the present floating position. The respective net heave force and trim moment values follow the flow-induced forces along the wetted boundaries $\Gamma^\mathrm{H}$ augmented by gravity forces, viz.
\begin{align}
F^\mathrm{ H} &= \frac{g_\mathrm{i}}{|g_\mathrm{k}|} \left[ \int_\mathrm{\Gamma^\mathrm{ H}} f_\mathrm{i} \, \mathrm{d} \Gamma + V^\mathrm{ ini} \, \rho^\mathrm{ b} \, g_\mathrm{i} \right] \qquad \mathrm{and} \qquad \label{equ:net_heave_force} \\
M^\mathrm{ T} &= \frac{\epsilon_\mathrm{ijk} \, g_\mathrm{j} \, v_\mathrm{k}^{\mathrm{bulk}}}{|\epsilon_\mathrm{ijk} \, g_\mathrm{j} \, v_\mathrm{k}^{\mathrm{bulk}}|}  \left[ \int_\mathrm{\Gamma^\mathrm{ H}} \epsilon_\mathrm{ilm} \big[ x_\mathrm{l} - x_\mathrm{l}^\mathrm{ r} \big] f_\mathrm{m} \, \mathrm{d} \Gamma + V^\mathrm{ ini} \, \rho^\mathrm{ b} \epsilon_\mathrm{ilm} \big[ x_\mathrm{l}^\mathrm{ g} - x_\mathrm{l}^\mathrm{ r} \big] g_\mathrm{m} \right] \label{equ:net_trim_moment} \, , 
\end{align}
where $f_\mathrm{i} = \left[ 2 \, \mu^\mathrm{ eff} \, S_\mathrm{ik} - p^\mathrm{ eff} \, \delta_\mathrm{ik} \right] n_\mathrm{k}$ and $v_\mathrm{k}^{\mathrm{bulk}}$ represent the surface specific fluid forces and the bulk velocity, respectively. Here $\epsilon_\mathrm{ijk}$ refers to the Levi-Civita-Symbol used to compute an outer vector product. Once the flow field and the forces on the hull converge, the deviation from the hydrostatic floatation is evaluated according to Eqn. (\ref{equ:adaptive_trim_sinkge_error}). Subsequently, under-relaxed corrections are superimposed by means of a displacement vector $d_\mathrm{i}^\mathrm{ H}$ along each interior boundary surface element
\begin{align}
d_\mathrm{i}^\mathrm{ H} = \left[ \Delta S_\mathrm{3} \, \delta_\mathrm{i3} + \Delta T_\mathrm{2} \, R_\mathrm{i2} \right] \omega^\mathrm{H} \, , 
\end{align}
where $R_\mathrm{i2}$ refers to the entries of a rotation matrix around the trim axis. Robust convergence was experienced for $0.2 \leq \omega^\mathrm{H} \leq 0.6$.

Mind that the Lagrangian (\ref{equ:simplify_derive_lagragian_rans}) could be augmented by residual versions of the floatation model (\ref{equ:net_heave_force}) and (\ref{equ:net_trim_moment}) to implicitly account for the influence of the floatation on the control using entries of the floatation to the adjoint PDE system (\ref{equ:adjoint_rans_mass})-(\ref{equ:adjoint_rans_momentum}). As the floatation updates are, however, usually fairly small and simulations only aim at steady state floatation, we do not consider this approach.

\subsection{Mesh Deformation Procedure}
\label{sec:meshdeform}
The interior boundary displacement and the fixed exterior boundaries serve as Dirichlet conditions for a mesh morphing routine which updates the interior cell centers from a CDS-based FV approximation of a Laplace equation 
\begin{align}
\frac{\partial}{\partial x_\mathrm{k}} \left[ \mu^\mathrm{ d} \frac{\partial d_\mathrm{i}}{\partial x_\mathrm{k}} \right]= 0 \quad &\mathrm{in} \quad \Omega
\qquad \mathrm{with} \qquad
\begin{cases} 
d_\mathrm{i} = d_\mathrm{i}^\mathrm{H} \quad &\mathrm{on} \quad \Gamma \cap \Gamma^\mathrm{ H} \\
d_\mathrm{i} = 0 \quad &\mathrm{on} \quad \Gamma 
\end{cases} \, . \label{equ:adaptive_trim_sinkage_field_equation}
\end{align}
In the present study, the diffusivity field $\mu^\mathrm{ d}$ refers to the inverse (non zero) distance to the nearest wall, which avoids a grid distortion in the vicinity of the hull. A subsequent deformation of the cell vertices follows from an averaged interpolation of all vertex-adjacent centers $N^\mathrm{P(V)}$, viz.
\begin{align}
d_\mathrm{i}^\mathrm{ V} = \frac{1}{N^\mathrm{P(V)}} \sum_\mathrm{P=1}^{N^\mathrm{P(V)}} \left[ d_\mathrm{i}^\mathrm{ P} + \frac{\partial \, d_\mathrm{i}^\mathrm{ P}}{\partial \, x_\mathrm{k}} \left( x_\mathrm{k}^\mathrm{ V} - x_\mathrm{k}^\mathrm{ P} \right) \right] \label{equ:mesh_update} \, .
\end{align}
After updating the grid vertices and CV centers, the geometric quantities are recalculated for each CV. Equations (\ref{equ:adaptive_trim_sinkage_field_equation}) and (\ref{equ:mesh_update}) are employed to update the volume grid in response to the change of the discrete hull, which in turn can alter on the basis of either the floating body motion ($d_\mathrm{i} = d_\mathrm{i}^\mathrm{H}$, Sec. \ref{subsec:floatation}) or the computed sensitivities $s$ of the optimization ($d_\mathrm{i} \sim s \, n_\mathrm{i}$, cf. Sec \ref{sec:shape_gradient}) that enter the mesh deformation approach through the boundary conditions. Different approaches to define the diffusivity field $\mu^\mathrm{ d}$ are conceivable, though we do consistently use an inverse distance-based approach in this study. Since the grid topology remains unaltered, the CFD simulation is continued from the previous result on the ''new'' mesh. 

The following sections \ref{sec:shape_gradient} and \ref{subsec:geometrical_constraints} outline the computation of a constraint compatible, cell center gradient field $g_\mathrm{i}$ derived from the adjoint sensitivities. Using a step size $\alpha^\mathrm{d}$ this is translated into an optimization based relocation of cell centers and subsequently fed into (\ref{equ:mesh_update}).

\subsection{Node-based Shape Gradient Approximation }
\label{sec:shape_gradient}
Non-parameterized, node-based shape optimizations disclose localized influences on optimal shapes down to the range of the discrete surface elements of the CFD grid. However, the strategy also suffers from a few well-known weaknesses. For example, the raw sensitivities provide comprehensive information on the normal deformation but lack any information on the associated tangential node motion, and the sensitivities are not necessarily smooth. These deficiencies yield rough/noisy shape updates, cf. \cite{stuck2011adjoint, kroger2015cad}, and lead to distorted near-wall meshes, which in turn hamper the preservation of numerical accuracy during the optimization procedure, e.g. \cite{stavropoulou2014plane} and \cite{bletzinger2014consistent}. 

The adjoint shape derivatives are usually regularized to obtain smooth meaningful technical shape updates. Different regularization strategies to determine the shape gradient exist, which 
can be distinguished by the surface- or volume-based habitat of the shape gradient. The most prominent example refers to a surface-based formulation using the Laplace-Beltrami (LB) metric, as initially proposed by \cite{jameson2000studies} and \cite{vassberg2006aerodynamic, vassberg2006aerodynamic_II} in terms of an implicit, continuous smoothing operator based on an extended definition of the inner product, frequently labeled Sobolev-gradient. More recently, a volume-based Steklov-Poincaré (SP) metric was suggested as an alternative, e.g. \cite{schulz2016computational, haubner2021continuous}, which offers algorithmic and procedural benefits and shares features with the traction method introduced by \cite{azegami1996domain, azegami2006smoothing}.

\subsubsection{Steklov-Poincar\'e Metric}
\label{subsubsec:steklov_poincare}
The SP approach refers to a novel strategy on an industrial level that employs an elliptic volume-based formulation where smoothed results are subsequently projected on the boundary. The procedure essentially  combines the 2D shape update with the 3D mesh update using the discrete CFD mesh sensitivities. The algorithm exclusively operates in the fluid domain and is thus  compatible with the CFD solver environment. Re-using standard high-performance-computing capable solver routines (assembling, solving, etc.) represents a major benefit of the SP procedure which refers to a standard Laplace-PDE to compute a gradient vector field $g_\mathrm{i}$
\begin{align}
\frac{\partial}{\partial x_\mathrm{k}} \left[ \mu^\mathrm{ g} \frac{\partial g_\mathrm{i}}{\partial x_\mathrm{k}} \right]= 0 \quad &\mathrm{in} \quad \Omega
\qquad \mathrm{with} \qquad
\begin{cases} 
\frac{\partial \, g_\mathrm{i}}{\partial \, n} = s \, n_\mathrm{i} \quad &\mathrm{on} \quad \Gamma^\mathrm{} \cap \Gamma^\mathrm{ D} \\
g_\mathrm{i} \, n_\mathrm{i} = 0, \frac{\partial \, g_\mathrm{i}}{\partial \, n} t_\mathrm{i} = 0 \quad &\mathrm{on} \quad \Gamma^\mathrm{ Symm} \\
g_\mathrm{i} = 0 \quad &\mathrm{on} \quad \Gamma^\mathrm{}
\end{cases} \, , \label{equ:steklov_poincare_field_equation}
\end{align}
The gradient field $g_\mathrm{i}$ is controlled by Neumann conditions that employ the raw sensitivity derivatives $s$. No boundary-based operations are necessary and modifications  of the boundary conditions  (\ref{equ:steklov_poincare_field_equation}) support an intuitive introduction of additional geometry related engineering constraints. Examples refer to fixed intersection lines along a symmetry plane via $g_\mathrm{i} \, n_\mathrm{i} = 0$ and $(\partial \, g_\mathrm{i} / \partial \, n) t_\mathrm{i} = 0$ on $\Gamma^\mathrm{Symm}$, or the realization of a mandatory flat ship transom obtained by $g_\mathrm{i} \, n_\mathrm{i} = 0$. The SP approach involves only a single user-defined parameter, i.e. the diffusivity $\mu^\mathrm{ g}$, which refers to the inverse (non zero) distance to the nearest wall in the present study, cf. Sec. \ref{sec:meshdeform}. The method is the starting point for more sophisticated p-Laplacian descent strategies that  employ a nonlinear diffusivity, e.g. $\mu^\mathrm{ g} = [(\partial \, g_\mathrm{i} / \partial \, x_\mathrm{k}) (\partial \, g_\mathrm{i} / \partial \, x_\mathrm{k})]^{(p-2)/2}$, cf. \cite{mueller2021novel}.

The SP approach is the preferred approach of this paper. A step in the steepest descent direction is performed once the field $g_\mathrm{i}$ is computed from (\ref{equ:steklov_poincare_field_equation}) and subsequently subjected to further technical constraints, cf. Sec. \ref{subsec:optimization_procedure}.

\subsection{Geometrical Constraints}
\label{subsec:geometrical_constraints}
The discussion of additional geometrical constraints is divided into local and global (integral) criteria.

\subsubsection{Local Constraints}
\label{subsubsec:local_geometrical_constraints}
Local constraints restrict the motion of the shape in Euclidean space. For example, marine engineering examples typically refer to a maximum length, a maximum width, or a plane transom stern. Various strategies are conceivable to meet local constraints. Superficially, all constraints can be incorporated on equation level to determine the field gradient, e.g. (\ref{equ:steklov_poincare_field_equation}). However, this essentially resembles a sub-optimization problem and --for performance reasons-- requires the availability of a suitable procedure, i.e., a Newton-type solver. 
Alternatively, augmented Lagrangian methods may be used, which relax the geometrical constraints by introducing additional Lagrangian multipliers, cf. \cite{allaire2004structural, andreani2008augmented, allaire2020geometric, mueller2021novel}. The latter serve as additional process parameters and usually result in more optimization cycles, especially if several geometrical constraints should be considered simultaneously. 

The present procedure augments the flow sensitivity to comply with local constraints, is modularizable, and intuitive to use. By reference to an exemplary geometric inequality that constrains the maximum control-coordinate $\tilde{u}_\mathrm{i}$ in $x_\mathrm{i}$-direction, the optimization problem is augmented 
\begin{align}
\mathrm{min} \, J(\varphi (u_\mathrm{i}), u_\mathrm{i}) \qquad \mathrm{s.t.} \qquad \mathrm{R}^\mathrm{ \varphi}(\varphi (u_\mathrm{i})) = 0 \qquad \mathrm{and} \qquad u_\mathrm{i} - \tilde{u}_\mathrm{i} \leq 0 \, ,
\end{align}
and the sensitivity of the shape w.r.t. the flow $s \, n_\mathrm{i}$ is assumed to be available. As long as the shape remains below the upper  bound $u_\mathrm{i} - \tilde{u}_\mathrm{i} \leq 0$, the  constraint is inherently fulfilled. However, if the shape moves beyond the boundary, an additional compensating geometric sensitivity $s_\mathrm{i}^\mathrm{ u}$ is added to the flow sensitivity, viz. 
\begin{align}
s \, n_\mathrm{i} \to s \, n_\mathrm{i} +  \beta^\mathrm{u} \, s_\mathrm{i}^\mathrm{ u}
\qquad \mathrm{with} \qquad 
s_\mathrm{i}^\mathrm{ u} = 
\begin{cases}
0 \quad &: u_\mathrm{i} - \tilde{u}_\mathrm{i} \leq 0 \\
\tilde{u}_\mathrm{i} - u_\mathrm{i}  \quad &: u_\mathrm{i} - \tilde{u}_\mathrm{i} > 0
\end{cases} \, . \label{equ:local_geometrical_constraint}
\end{align}
Compliance of dimensions and scaling of the geometric constraint is ensured by appropriate choices of the constant $\beta^\mathrm{ u }$. A natural choice of the scaling refers to the inverse step size of the employed steepest descent approach, i.e., $\beta^\mathrm{ u } =1/ \alpha^\mathrm{d}$.

\subsubsection{Global Constraints} 
\label{subsubsec:global_geometrical_constraints}
Global constraints require the preservation of integral quantities such as, e.g., the hydrostatic water displacement, a maximum wetted surface, or a fixed center of gravity. The displacement is particularly important in this work since drag optimizations often tend to eliminate the wetted surface and the hull. 

Implicit SP procedures directly employ a field equation (\ref{equ:steklov_poincare_field_equation}) using the sensitivity derivatives as Neumann conditions. This suggests solving an analog, volume-based sub-problem to preserve global constraints, e.g., plane transom surfaces or fixed mainframes, viz. 
\begin{align}
\frac{\partial}{\partial x_\mathrm{k}} \left[ \mu^\mathrm{g} \frac{\partial \tilde{g}_\mathrm{i}}{\partial x_\mathrm{k}} \right]= 0 \quad &\mathrm{in} \quad \Omega
\qquad \mathrm{with} \qquad
\begin{cases} 
\frac{\partial \, \tilde{g}_\mathrm{i}}{\partial \, n} = n_\mathrm{i} \quad &\mathrm{on} \quad \Gamma^\mathrm{} \cap \Gamma^\mathrm{ D} \\
\tilde{g}_\mathrm{i} \, n_\mathrm{i} = 0, \frac{\partial \, \tilde{g}_\mathrm{i}}{\partial \, n} t_\mathrm{i} = 0 \quad &\mathrm{on} \quad \Gamma^\mathrm{Symm} \\
\tilde{g}_\mathrm{i} = 0 \quad &\mathrm{on} \quad \Gamma
\end{cases} \, . \label{equ:adjoint_steklov_poincare_field_equation}
\end{align}
The approach shares ideas of a sub-optimization problem that aims minimizing the squared integral of the deformation flux through the wetted part of the shape constraint by the SP field gradient Eqn. (\ref{equ:steklov_poincare_field_equation}), viz. $ J^\mathrm{ \Gamma, D} = \left[ \int_\mathrm{\Gamma^\mathrm{ W}} g_\mathrm{i} \, n_\mathrm{i} \, \mathrm{d} \Gamma \right]^2$. Following the numerical solution of (\ref{equ:adjoint_steklov_poincare_field_equation}) a long the lines of (\ref{equ:steklov_poincare_field_equation}), a superposition of the $g_\mathrm{i}$ and $\tilde{g}_\mathrm{i}$ fields yields a conservative volume-based shape gradient, viz.
\begin{align}
g_\mathrm{i} 
%\gets 
\to g_\mathrm{i} + 
\beta^\mathrm{ g } \, \tilde{g}_\mathrm{i}
\qquad \mathrm{with} \qquad
\beta^\mathrm{ g} = -\frac{ \int_\mathrm{\Gamma^\mathrm{ W}} g_\mathrm{i} \, n_\mathrm{i} \, \mathrm{g} \Gamma}{ \int_\mathrm{\Gamma^\mathrm{ W}} \tilde{g}_\mathrm{i} \, n_\mathrm{i} \, \mathrm{d} \Gamma} \, .
\end{align}
After a final scaling of the deformation field with a negative step size $d_\mathrm{i} \to - \alpha^\mathrm{d} g_\mathrm{i}$, the cell centered deformation field is used to adjust the grid vertices from (\ref{equ:mesh_update}). The explicit LB approach is used for validation purposes in Sec. \ref{sec:validation}. All applications in Sec. \ref{sec:application} employ the SP approach, i.e. approximate (\ref{equ:steklov_poincare_field_equation}) as well as (\ref{equ:adjoint_steklov_poincare_field_equation}).

\section{Numerical Procedure}
\label{sec:Numerik}
The numerical procedure utilizes a Finite-Volume (FV) approximation, dedicated to Single Instruction Multiple Data (SIMD) implementations on a distributed-memory parallel CPU machine. Algorithms employed by the inhouse procedure FresCo$^+$ are described in \cite{rung2009challenges}, and \cite{yakubov2013hybrid}. They ground on the integral form of a generic Eulerian transport equation for a scalar field $\phi(x_\mathrm{k}, t)$ exposed to the influence of a possibly non-linear source term $S_\phi$ in addition to a modeled (non-linear) gradient diffusion $\Gamma^*$ in a control volume $V$ bounded by the Surface $S(V)$, viz. 
\begin{align}
\label{Momentum}
\int_V \left[ \frac{\partial \phi}{\partial t} - S_\phi \right] dV +\oint_{S(V)} dS_\mathrm{i} \left[  v_\mathrm{i} \phi -  
  (\Gamma + \Gamma^*) \frac{\partial \phi}{\partial x_\mathrm{i}} \right] = 0 \; . 
\end{align}
The procedure uses the strong conservation form and employs a cell-centered, co-located storage arrangement for all transport properties. The spatial discretization employs unstructured grids based on arbitrary polyhedral cells, which connect to a face-based data structure. Various turbulence-closure models are available w.r.t. statistical (RANS) or scale-resolving (LES, DES) approaches. The numerical integration refers to the mid-point rule, diffusive fluxes are determined from second-order central differencing, and convective fluxes employ higher-order upwind biased interpolation formulae. Preconditioned Krylov-subspace solvers are used to solve the equation systems, and the global flow field is iterated to convergence using a pressure-correction scheme. Procedures are parallelized using a domain decomposition method and the MPI communication protocol.

\subsection{Optimization Procedure}
\label{subsec:optimization_procedure}
After all engineering constraints are incorporated, either locally on sensitivity level or globally within the shape gradient computation, a descent procedure is employed to minimize the cost functional. For this purpose, the volume-based representation of the shape gradient is multiplied by a sufficiently small step size $d_\mathrm{i} \to -\alpha^\mathrm{d} \, d_\mathrm{i}$ [$d_\mathrm{i} = -\alpha^\mathrm{d} \, g_\mathrm{i}$]  for the SP [LB] metric that a) ensures compliance of dimensions between the LB or SP-based shape gradient and b) serves as an optimization step in the direction of steepest descent. The step size remains constant over the optimization process and is frequently estimated based on a maximum initial displacement, i.e. $\alpha^\mathrm{d} = d^\mathrm{max} / \mathrm{max}(d_\mathrm{i}, g_\mathrm{i})$. Typical values for this maximum displacement refer to $1/10^{4} \leq d^\mathrm{max} / L \leq 1/10^{3}$, where $L$ denotes to a reference length of the underlying geometry, e.g. the ship length. 
Subsequent deformation of the cell vertices follows from an averaged interpolation of all vertex-adjacent centers $N^\mathrm{P(V)}$ in line with Eqn. (\ref{equ:mesh_update}).
After updating the grid, geometric quantities are recalculated for each CV. Topological relationships remain unaltered, and the simulation is continued by a restart from the previous optimization step to evaluate the new objective functional value. Due to the employed steepest descent approach and comparably small step sizes, field solutions of two consecutive shapes are usually nearby. Compared to a simulation from scratch, a speedup in total computational time of about an order of magnitude is realistic for this papers' applications.
The optimization loop is terminated if a maximum number of optimization cycles $N^\mathrm{O}$ is reached or if the relative cost functional decrease w.r.t. the initial shape falls below $\epsilon^\mathrm{J} [\%]$ during an optimization step, cf. Alg. \ref{alg:optimization_procedure}.
\begin{algorithm}
define: $\mathrm{d}^\mathrm{max}$, $\mathrm{N}^\mathrm{O}$ and $\epsilon^\mathrm{J}$ \\
$n^\mathrm{opt} = 1$ \\
\While{($n^\mathrm{opt} \leq \mathrm{N}^\mathrm{O}$)}{
approximate primal two-phase system, cf. (\ref{equ:primal_rans_mass})-(\ref{equ:primal_rans_momentum}) \\
evaluate cost functional $J$ \\
\eIf{($n^\mathrm{opt} > 1$) $\mathrm{and}$ $(J - J^\mathrm{ ini})/J^\mathrm{ ini} \cdot 100 \leq \epsilon^\mathrm{J}$)}
{\vspace{0.1cm}
\textbf{terminate}
}
{
approximate adjoint two-phase system, cf. (\ref{equ:adjoint_rans_mass})-(\ref{equ:adjoint_rans_momentum}) \\
compute shape (sensitivity) derivative w.r.t. the fluid flow $s$, cf. (\ref{equ:shape_derivative}) \\
employ local geometric constraint(s) and manipulate shape derivative, e.g. (\ref{equ:local_geometrical_constraint}) \\
approximate shape gradient
%, e.g. (\ref{equ:laplace_beltrami}) or 
(\ref{equ:steklov_poincare_field_equation}) \\
employ global geometric constraint(s) and manipulate the shape gradient
%, e.g. (\ref{equ:gradient_projection}) or
(\ref{equ:adjoint_steklov_poincare_field_equation}) \\
define: $d_\mathrm{i} = -\alpha^\mathrm{d} \, g_\mathrm{i}$\\
perform a domain (shape) update (\ref{equ:mesh_update})
}
$n^\mathrm{opt} \to n^\mathrm{opt} + 1 $
}
\caption{Schematic representation of the employed gradient descent procedure, where $N^\mathrm{O}$, $\mathrm{d}^\mathrm{max}$, $\epsilon^\mathrm{J}$, and $J^\mathrm{ ini}$ denote the maximum optimization iteration, a user-defined maximum deformation, the objective convergence criterion, and the initial ($n^\mathrm{opt} = 1$) cost functional value, respectively.}
\label{alg:optimization_procedure}
\end{algorithm}

%% file: tex/validation.tex
\begin{figure}[!ht]
\centering
\subfigure[]{
\iftoggle{tikzExternal}{
\input{./tikz/cylinder_scetch.tikz}}{
\includegraphics{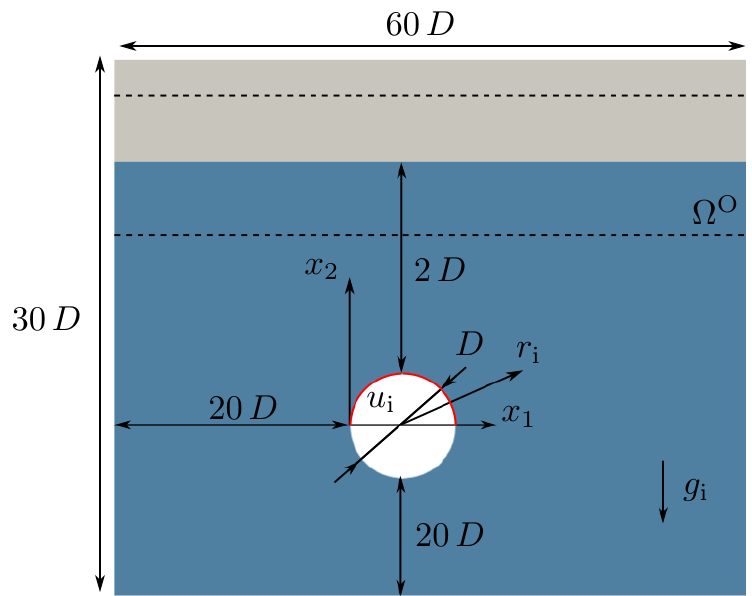}}
}
\subfigure[]{
\includegraphics[scale=1]{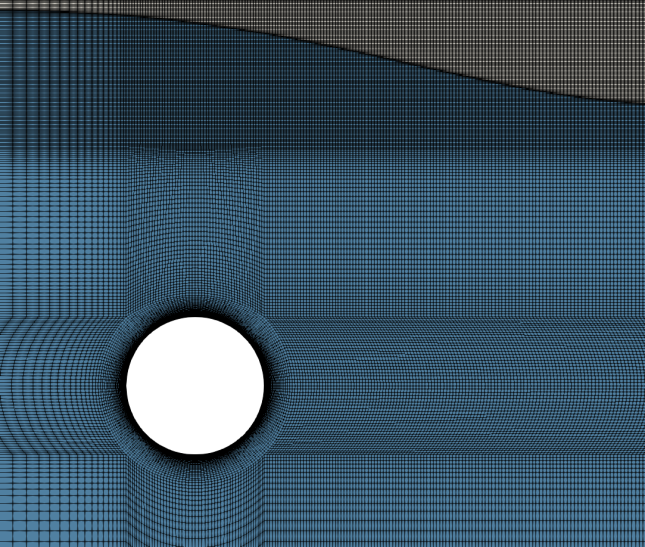}
}
\caption{Illustration of the submerged 2D laminar cylinder validation case ($\mathrm{Re}_\mathrm{D} = 20$, $\mathrm{Fn}=0.75$): (a) Schematic drawing of the initial configuration including the controlled upper half of the cylinder $u_\mathrm{i}$ (red) and (b) structured-grid portion in the vicinity of the cylinder.}
\label{fig:cylinder_fn_075}
\end{figure}

This section assesses the credibility of the adjoint two-phase flow sensitivities against the results of a Finite Difference (FD) approach. Additionally, influences of the adjoint two-phase flow couplings are investigated. The considered example refers to the laminar flow around a two-dimensional submerged circular cylinder at fixed floatation and involves volume- and surface-based cost functionals. As illustrated in Fig. \ref{fig:cylinder_fn_075} (a), the origin of the cylinder is positioned two and a half diameters $D$ underneath an initial calm-water free surface. The employed two-dimensional domain features a length and height of $60 \, D$ and $30 \, D$, where the inlet and bottom boundaries are located 20,5 diameters away from the cylinder's origin. At the inlet, a homogeneous unidirectional  (horizontal) bulk flow $v_\mathrm{i} = v_\mathrm{1} \delta_{i1}$ is imposed for both phases in conjunction with a calm water concentration distribution. Slip walls are used along the top and bottom boundaries, and a hydrostatic pressure boundary is employed along with the outlet. The grid is stretched in the longitudinal direction ($x_\mathrm{1}$) towards the outlet to suppress the outlet wave field and comply with the outlet condition.
  
The study is performed at $\mathrm{Re}_\mathrm{D} = v_\mathrm{1} D/\nu^\mathrm{ b} = \SI{20}{}$ and $\mathrm{Fn} = v_\mathrm{1}/\sqrt{G \, 2 \, D} = \SI{0.75}{}$, based on the gravitational acceleration $G$, the inflow velocity $v_\mathrm{1}$ and the kinematic viscosity of the water $\nu^\mathrm{ b}$. The expected dimensionless wave length reads $\lambda = \lambda/ D = 2 \, \pi \, \mathrm{Fn}^2 = 3.534$. To ensure the independence of the objective functional value w.r.t. spatial discretization, a grid study was conducted prior to the optimization study. Part of the utilized structured numerical grid is displayed in Fig. \ref{fig:cylinder_fn_075} (b). It consists of approximately $\SI{215000}{}$ control volumes where the cylinder shape is discretized with 500 surface elements along the circumference. 
The non-dimensional wall-normal distance of the first grid layer reads $y^+ \approx \SI{0.01}{}$ and the refined grid in the free surface region employs isotropic spacing with $\Delta x_\mathrm{1} =\Delta x_\mathrm{2} \approx \lambda/100$. Convective primal [adjoint] momentum fluxes are approximated using the QUICK [QDICK] scheme, cf. \cite{stuck2013adjoint}. The approximation of the concentration equation has been outlined in \cite{kuhl2021adjoint, kuhl2021cahn}, where traditional VoF approaches follow from a  compressive primal/hybridized continuous-discrete adjoint HRIC scheme and compressiveness of a CH-VoF is achieved through inherent phase separation capabilities outlined in Sec \ref{sec:conctrans}. Using an Euler implicit approach, the simulations are advanced to a steady state in pseudo time.

\subsection{Local Validation}
\label{sec:val-local}
Three exemplary objectives  are used to validate the adjoint two-phase model against FD results at selected positions along the circumference of the cylinder.  Results are reported for two boundary-based force objectives into the direction $r_\mathrm{i} = [\sqrt(2), \sqrt{2}]^\mathrm{T} / 2$  (drift) and  $r_\mathrm{i} = [1, 0]^\mathrm{T}$ (drag), and a volumetric target-concentration objective with a habitat along $\Omega^\mathrm{ O} = [-5D,D] \times [25 D,5 D]$. The credibility of the validation effort is ensured by verifying the linearity of the FD-analysis using three perturbation magnitudes  $\epsilon / D \in [10^{-4}, 10^{-5}, 10^{-6} ] $. The  control is  restricted to the upper half of the cylinder (cf. Fig \ref{fig:cylinder_fn_075}a), for which FD results are extracted at 21 discrete positions. 
To this end, 42 additional simulations were performed to obtain second-order accurate central differences. 

\begin{figure}[!ht]
\centering
\iftoggle{tikzExternal}{
\input{./tikz/cylinder_fd_drift_inverse.tikz}
}{
\includegraphics{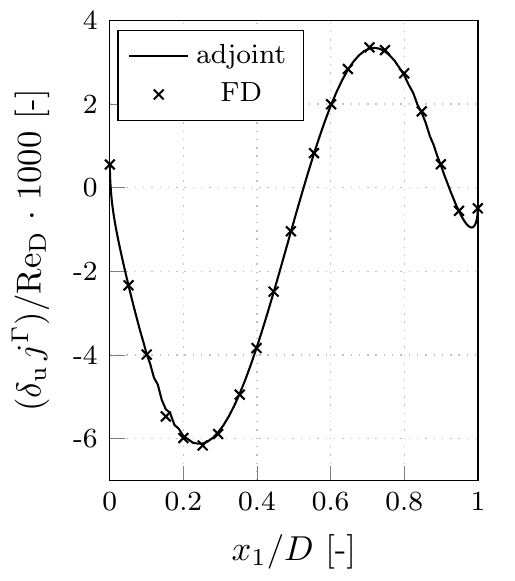}
\includegraphics{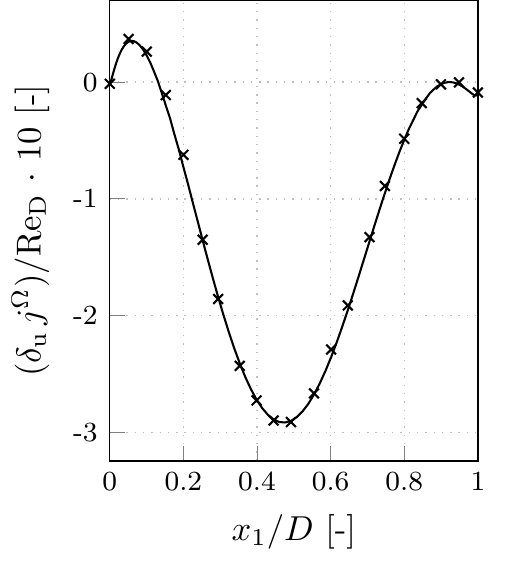}
\includegraphics{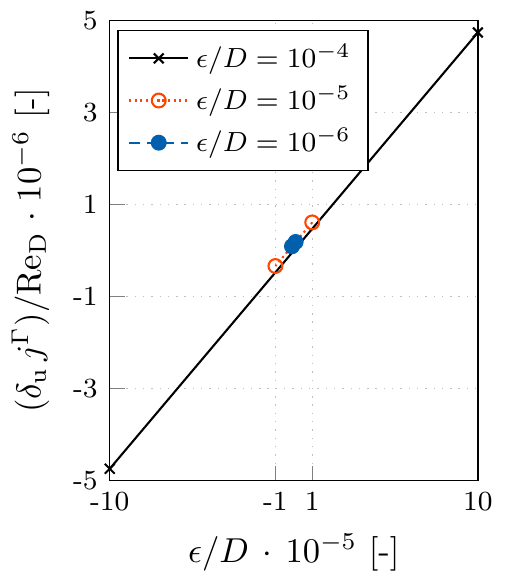}
}
\caption{Submerged cylinder validation case ($\mathrm{Re}_\mathrm{D} = 20$, $\mathrm{Fn}=0.75$): Continuous as well as discrete finite-difference (FD, $\epsilon / D =10^{-5}$) based sensitivity derivative along the upper cylinder side for (left) a drift functional ($r_\mathrm{i} = [\sqrt(2), \sqrt{2}]^\mathrm{T} / 2$), (center) the target concentration objective ($\Omega^\mathrm{ O} = [-5D,D] \times [25 D,5 D]$) as well as (right) three exemplary finite (force functional) system answers at $x_\mathrm{1} / D = 1/4$.}
\label{fig:cylinder_fd_study}
\end{figure}

A comparison of the sensitivities predicted by the adjoint (lines) and the FD (symbols) approaches is depicted in Fig. \ref{fig:cylinder_fd_study} for the drift objective (left) and the concentration objective (center) using $\epsilon / D =10^{-5}$. As indicated by these figures, the adjoint sensitivities agree almost perfectly with the FD results. The linearity of the FD answer is demonstrated in Fig. \ref{fig:cylinder_fd_study} (right), which refers to the local sensitivities for the force objective at an exemplary surface position $x_\mathrm{1} / D = 1/4$. Mind that the drift force objective deliberately promotes an interaction between the adjoint velocity and the gravity vector through the third RHS-term in (\ref{equ:adjoint_rans_concentration}) near the wall, where the boundary condition requires $\hat v_\mathrm{i} = - r_\mathrm{i}$  and therefore $\hat v_\mathrm{i} \, g_\mathrm{i} \neq 0$.

To analyze the sensitivity influence, especially from the individual contributions of the concentration Eqn. (\ref{equ:adjoint_rans_concentration}) via selective term combinations, nine additional adjoint studies were performed, cf. Tab.  \ref{tab:submerged_cylinder_systems}. Formulations A1-A3 neglect either all four source terms (A1), only the adjoint transposed convection (ATC) term due to the nonlinear momentum convection (A2), or all coupling terms induced by the differentiation of the material properties (A3), respectively. The deficit of neglecting different property-change sources in combination with the ATC term is investigated in A4-A6. The benefit of the individual source terms related to the change of the properties is in the focus of A7-A9.
\begin{table}[!ht]
\begin{center}
\begin{tabular}{|c||c|c|c|c|}
\hline
%$S^\mathrm{\hat{\varphi}}$ & 
%$\hat{v}_\mathrm{k} \frac{\partial \, v_\mathrm{k}}{\partial \, x_\mathrm{i}}$ &
%$\rho^\mathrm{ \Delta} \, \hat{v}_\mathrm{i} \, v_\mathrm{k} \frac{\partial \, %v_\mathrm{i}}{\partial \, x_\mathrm{k}}$ &
%$\frac{2 \, \mu^\mathrm{ \Delta}}{\mathrm{Re}}  S_\mathrm{ik} \frac{\partial \, %\hat{v}_\mathrm{i}}{\partial \, x_\mathrm{k}}$ & 
%$\frac{\rho^\mathrm{ \Delta} } {\mathrm{Fn}^2}  \hat{v}_\mathrm{i} \, g_\mathrm{i}$ \\
%\hline
%\hline
$S^\mathrm{\hat{\varphi}}$ & 
$\hat{v}_\mathrm{k} \frac{\partial \, v_\mathrm{k}}{\partial \, x_\mathrm{i}}$ &
$\rho^\mathrm{ \Delta} \, \hat{v}_\mathrm{i} \, v_\mathrm{k} \frac{\partial \, v_\mathrm{i}}{\partial \, x_\mathrm{k}}$ &
$2 \, \mu^\mathrm{ \Delta}  S_\mathrm{ik} \frac{\partial \, \hat{v}_\mathrm{i}}{\partial \, x_\mathrm{k}}$ & 
$\rho^\mathrm{ \Delta}  \hat{v}_\mathrm{i} \, g_\mathrm{i}$ \\
\hline
\hline
Label &ATC & convective term & Reynolds term & Froude term \\
\hline
\hline
A1 & - & - & - & - \\
\hline
A2 & - & x & x & x \\
\hline
A3 & x & - & - & - \\
\hline
\hline
A4 & x & - & x & x \\
\hline
A5 & x & x & - & x \\
\hline
A6 & x & x & x & - \\
\hline
\hline
A7 & - & x & - & - \\
\hline
A8 & - & - & x & - \\
\hline
A9 & - & - & - & x \\
\hline
\end{tabular}
\end{center}
\caption{Investigated adjoint source term configurations for the submerged cylinder validation case ($\mathrm{Re}_\mathrm{D} = 20$, $\mathrm{Fn}=0.75$), where '-' indicates a neglect of the respective contribution.}
\label{tab:submerged_cylinder_systems}
\end{table}

\begin{figure}[!ht]
\centering
\iftoggle{tikzExternal}{
\input{./tikz/cylinder_local_drag.tikz}
}{
\includegraphics{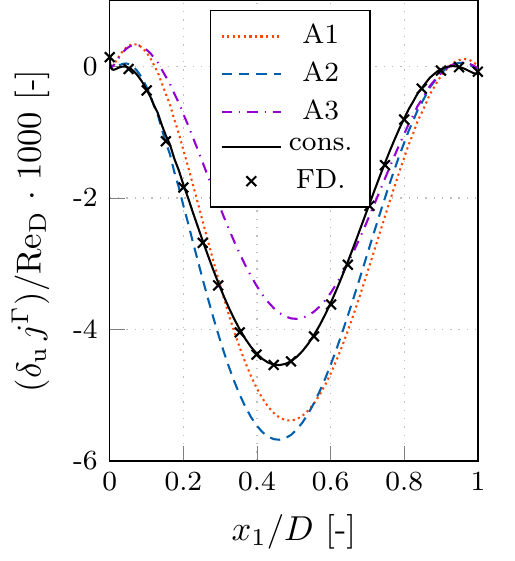}
\includegraphics{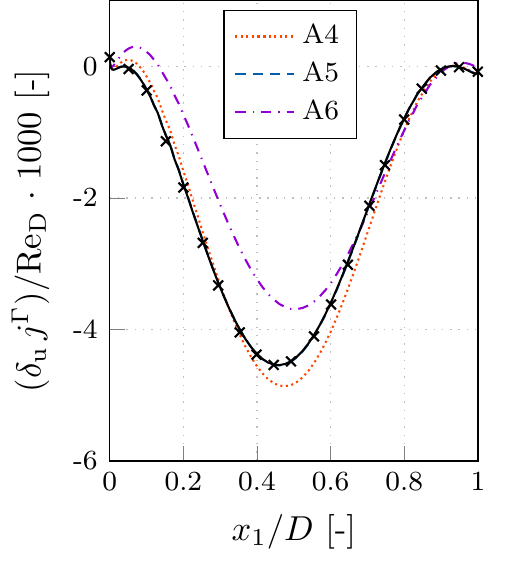}
\includegraphics{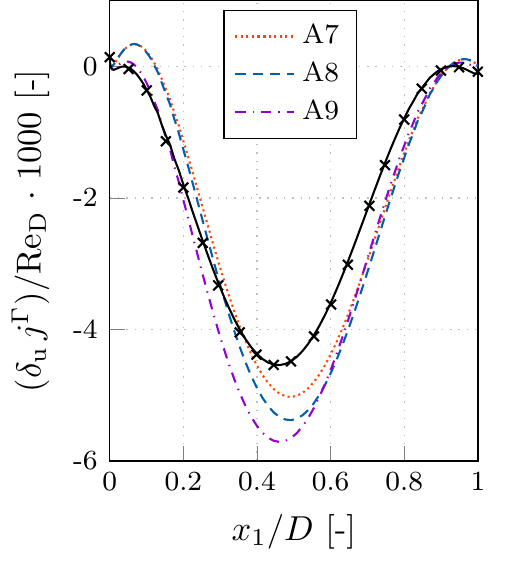}
}
\caption{Submerged cylinder validation case ($\mathrm{Re}_\mathrm{D} = 20$, $\mathrm{Fn}=0.75$): Continuous sensitivity derivative along the upper cylinder side using a drag functional ($r_\mathrm{i} = [1, 0]^\mathrm{T}$) for different adjoint systems A1-A9,
%that break the dual consistency via a modification of adjoint sources,
cf. Tab. \ref{tab:submerged_cylinder_systems}.}
\label{fig:cylinder_local_drag}
\end{figure}

\begin{figure}[!ht]
\centering
\iftoggle{tikzExternal}{
\input{./tikz/cylinder_local_inverse.tikz}
}{
\includegraphics{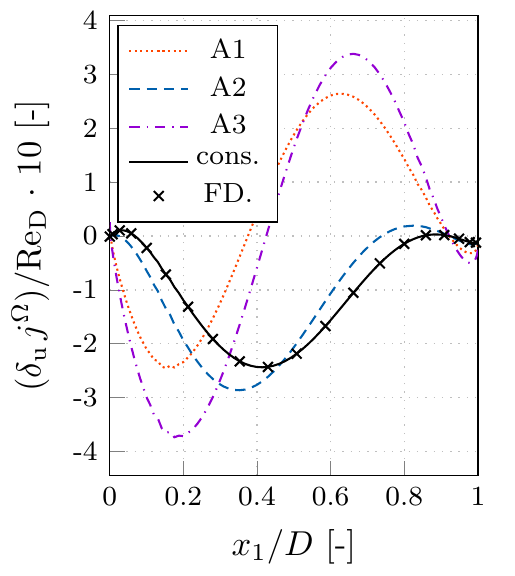}
\includegraphics{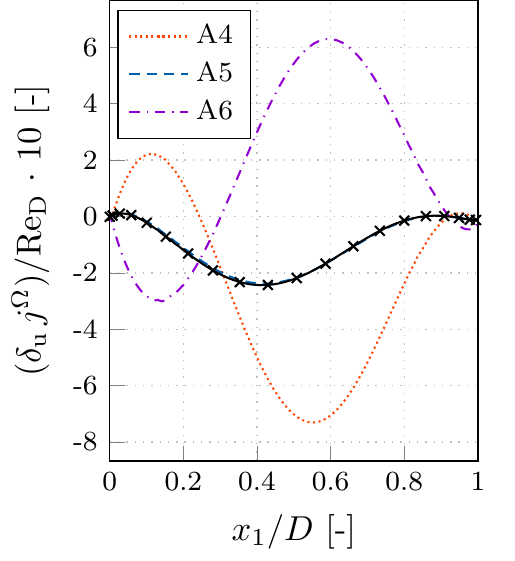}
\includegraphics{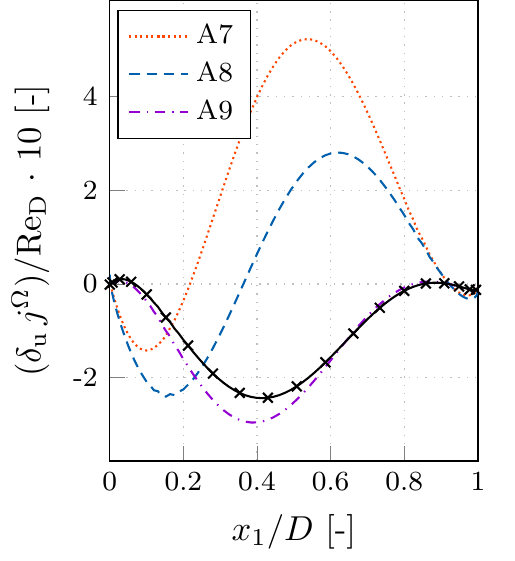}
}
\caption{Submerged cylinder validation case ($\mathrm{Re}_\mathrm{D} = 20$, $\mathrm{Fn}=0.75$): Continuous sensitivity derivative along the upper cylinder side using an inverse concentration objective ($\Omega_\mathrm{O} = [-5D,D] \times [25 D,5 D]$) for different adjoint systems A1-A9, 
%that break the dual consistency via a modification of adjoint sources, 
cf. Tab. \ref{tab:submerged_cylinder_systems}.}
\label{fig:cylinder_local_inverse}
\end{figure}

Results are displayed in Figs. \ref{fig:cylinder_local_drag} and \ref{fig:cylinder_local_inverse} for the drag force and the inverse concentration objective, respectively. Comparing A2 with A1 and A3 for the drag objective (Fig. \ref{fig:cylinder_local_drag}, left) reveals that ignoring convective, Reynolds, and Froude terms yields sign errors disregarding the ATC choice, which might be considered critical for gradient-based optimization. Significantly more pronounced sign errors are also observed for the concentration objective (Fig. \ref{fig:cylinder_local_inverse}, left) in conjunction with A1 and A3. The influence of the ATC term w.r.t. the total resistance objective resembles an overall sound influence while maintaining the qualitative characteristics. In line with the flat plate boundary-layer study from \cite{kuhl2021continuous}, an amplification of the shape derivative is obtained in its most sensitive region if the ATC term is neglected, which in turn can be treated based on reduced step sizes within a steepest descent optimization procedure. The situation becomes more crucial in the case of the inverse concentration objective. Already the neglect of the ATC term (A2) shifts the roots of the shape derivative. The manipulation of the sensitivity is significantly increased by neglecting the adjoint concentration sources (A3). 

A more detailed insight into the influence of the adjoint concentration sources is obtained by freezing selected terms (A4-A6). The resulting shape sensitivities are depicted in the central figures. The sensitivity deviations from the consistent formulation appear to be most significant when the contributions due to a variation of the density are neglected as only A5 reveals no sign errors. The variation of the Froude term seems to have the most extensive influence since A5 outperforms A4, which in turn improves on the results of A6. While a quantitative shift is observed for the surface-based functional, the deviations w.r.t. the volume-based cost functional are noticeably increased. Mind also the respective sign errors revealed by the center graph of Fig.  \ref{fig:cylinder_local_inverse}.

Finally, configurations A4-A6 are reversed by neglecting all except one source in A7-A9, cf. right graphs of Fig. \ref{fig:cylinder_local_drag} and Fig. \ref{fig:cylinder_local_inverse}. Considering only the Froude term (A9) underlines its major relevance by driving the shape sensitivity of the inverse concentration objective comparably close to the consistent result or towards the results of A2. Moreover, all but the solution for A9 render critical sign errors.

\subsection{Global Validation}
\label{subsec:global_validation}
Results presented in  Sec. \ref{sec:val-local} demonstrate a fair agreement between adjoint-based sensitivities and FD results, provided that a consistent formulation is employed. On the other hand, manipulations based on the neglect of adjoint coupling terms reveal both qualitative and quantitative influences on the sensitivities. The below reported global validation compares the convergence of  $(J - J^\mathrm{ ini}) / J^\mathrm{ ini}$ for different adjoint systems.  To this end, consistency influences on a complete optimization are assessed for ten formulations using either the consistent approach or one of the cases mentioned in Tab. \ref{tab:submerged_cylinder_systems} (A1-A9). The displacement of the cylinder was conserved during the optimization, and a constant maximum displacement $d^\mathrm{max} = D / 50$ was used to advance the shape. 

\begin{figure}[!ht]
\centering
\iftoggle{tikzExternal}{
\input{./tikz/cylinder_global_drag.tikz}
}{
\includegraphics{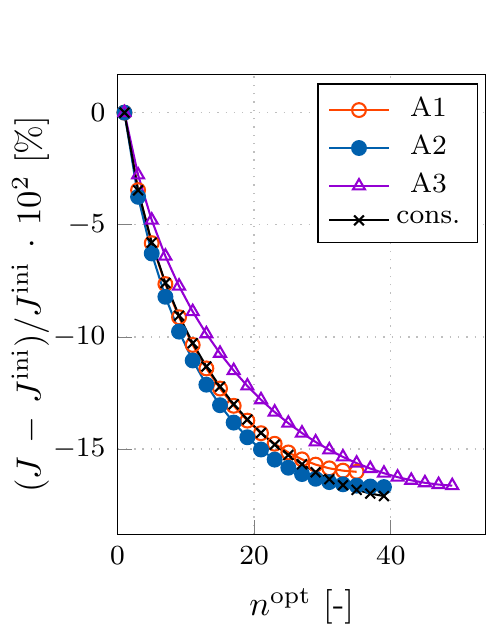}
\includegraphics{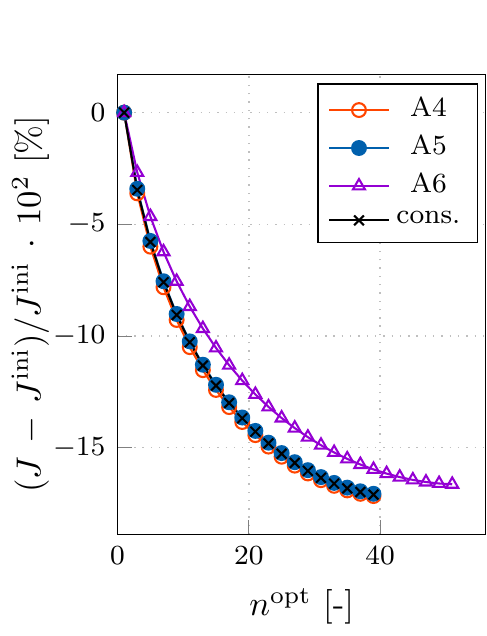}
\includegraphics{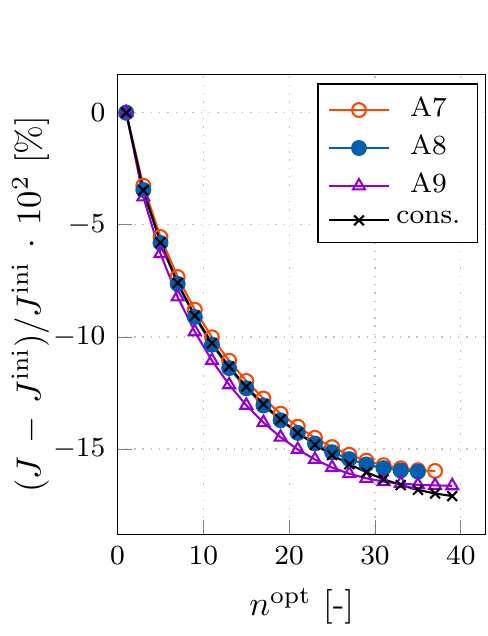}
}
\caption{Submerged cylinder validation case ($\mathrm{Re}_\mathrm{D} = 20$, $\mathrm{Fn}=0.75$): Convergence of the drag objective ($r_\mathrm{i} = [1, 0]^\mathrm{T}$) for different adjoint systems A1-A9,
%that break the dual consistency via a modification of adjoint sources,
cf. Tab. \ref{tab:submerged_cylinder_systems}.}
\label{fig:cylinder_global_drag}
\end{figure}

\begin{figure}[!ht]
\centering
\iftoggle{tikzExternal}{
\input{./tikz/cylinder_global_inverse.tikz}
}{
\includegraphics{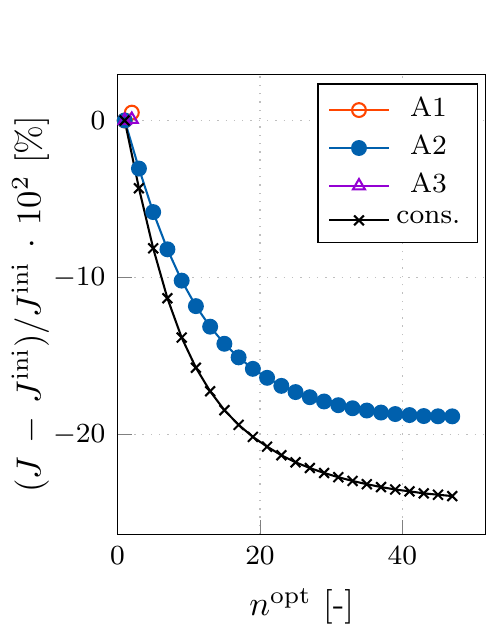}
\includegraphics{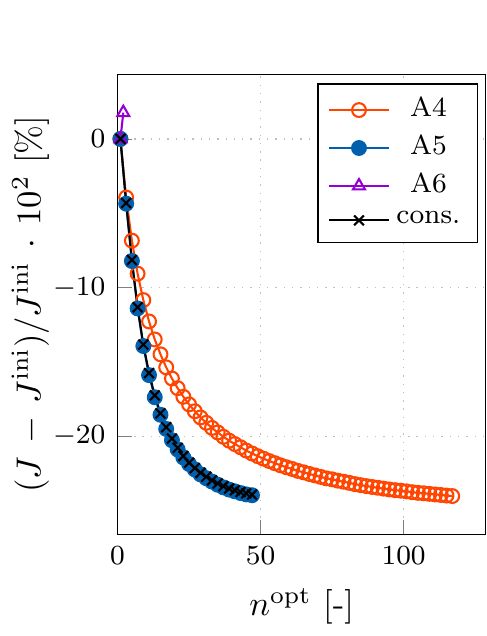}
\includegraphics{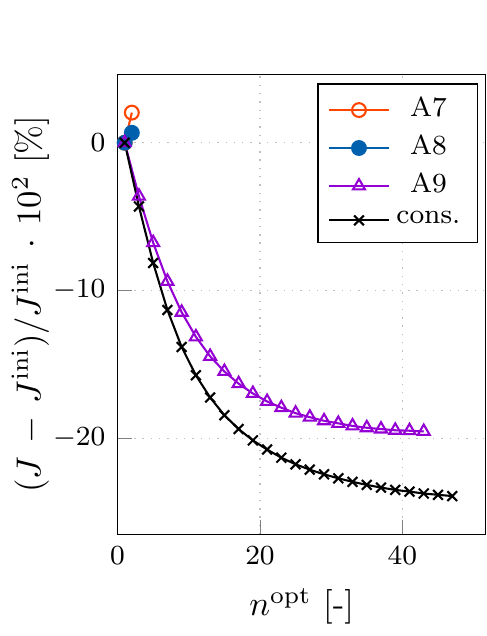}
}
\caption{Submerged cylinder validation case ($\mathrm{Re}_\mathrm{D} = 20$, $\mathrm{Fn}=0.75$): Convergence of the inverse concentration objective ($\Omega^\mathrm{ O} = [-5D,D] \times [25 D,5 D]$) for different adjoint systems A1-A9,
%that break the dual consistency via a modification of adjoint sources, 
cf. Tab. \ref{tab:submerged_cylinder_systems}.}
\label{fig:cylinder_global_inverse}
\end{figure}

Displayed results refer to the evolution of the normalized drag and the concentration objective as a function of the design candidate $n^{\mathrm{opt}}$. Figure \ref{fig:cylinder_global_drag} reveals that the various adjoint formulations return similar final drag values. The optimization gain w.r.t. drag is maximized if the fully consistent adjoint formulation is employed. However, neglecting all four adjoint source terms (A1) decreases the gain by $\approx 1\%$ only. The difference mostly arises within the last 20-30\% of the optimization, and the initial reduction of the cost functional is often similar. Exceptions refer to results for A3 and A6, which both neglect the  Froude term and underpin the relevance of the Froude term. When attention is directed to results obtained for the concentration-based objective depicted in Fig. \ref{fig:cylinder_global_inverse}, a more pronounced sensitivity to the formulation is observed. All adjoint systems that neglect the Froude term (A1-A2, A6-A8) perform a step into an ascent direction, and the optimization algorithm terminates. Moreover, the influence of the variation of the molecular viscosity (A5) appears to have a negligible impact.

Although the transfer of the results to turbulent marine engineering flows should be handled with caution, the validation results are indicative, particularly regarding the Froude term's importance.


%% file: tex/application.tex
\begin{figure}[!ht]
\centering
\includegraphics[scale=0.75]{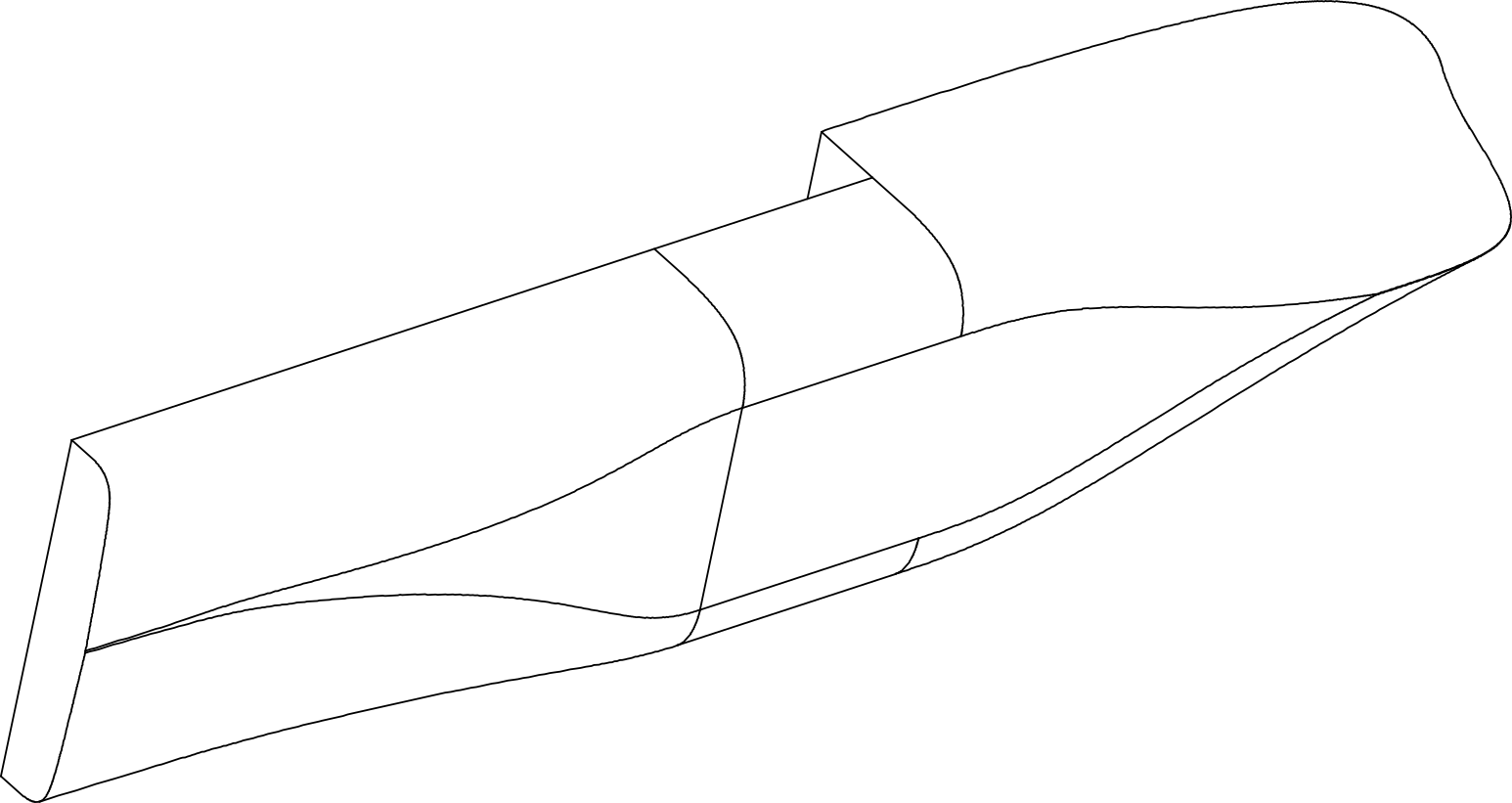}
\caption{Perspective 3D representation of the initial Offshore Supply Vessel (OSV). Mind that the geometry used by the numerical studies employs a simplified deck model, 
cf. Fig.  \ref{fig:osv_general}.}
\label{fig:osv_schick}
\end{figure}

The application study refers to an Offshore Supply Vessel (OSV, Fig. \ref{fig:osv_schick}) in full-scale (FS). While such vessels frequently cruise at large speeds, their hull length is often small. Therefore an OSV typically operates at large Froude numbers ($\mathrm{Fn} > 0.3$) and experiences large wave resistances based on, e.g., breaking waves. Minor modifications of the wave pattern might change the drag and substantially trigger a change of the floatation position. Thus the OSV case represents a challenging example for the present adjoint two-phase optimization framework under free floatation. Moreover, such vessels typically feature geometric constraints that render the quality of the constraint management method, the related descent strategy, and the capabilities of the mesh deformation procedure. 

\begin{figure}[!ht]
\centering
\subfigure[]{
\centering
\iftoggle{tikzExternal}{
\input{./tikz/osv_scetch.tikz}
}{
\includegraphics{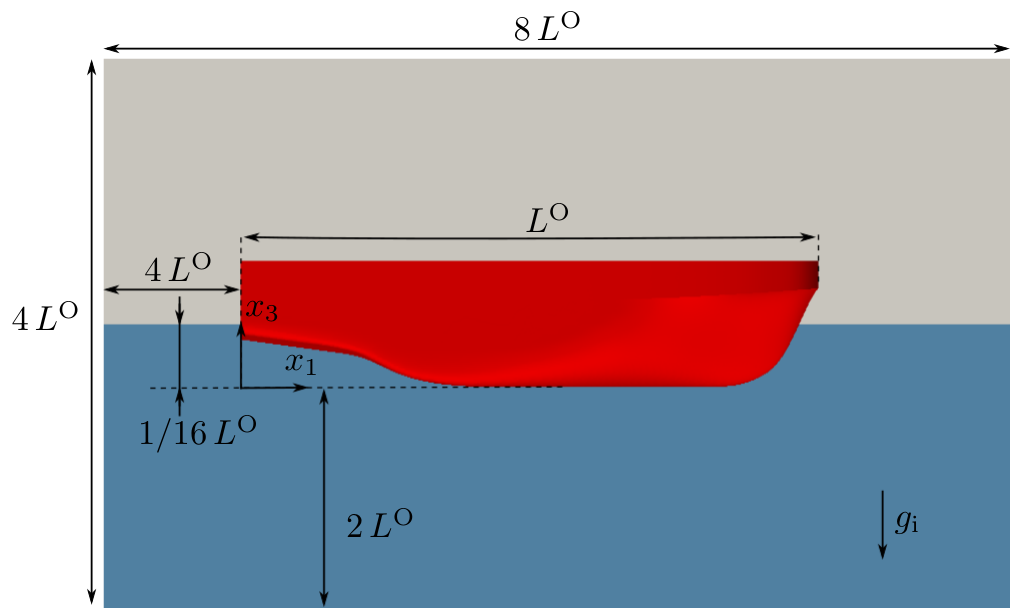}
}
}
\subfigure[]{
\centering
\includegraphics[scale=0.29]{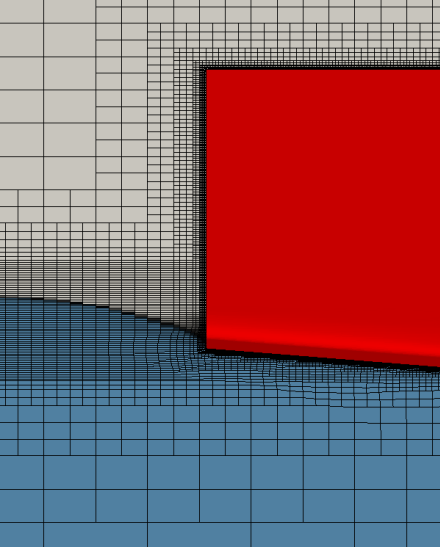}
}
\caption{ (a) Schematic drawing of the initial configuration and (b) detail of the unstructured  grid around the stern indicating  the free surface behind the full scale Offshore Supply Vessel (OSV).}
\label{fig:osv_general}
\end{figure}

\begin{figure}[!ht]
\centering
\includegraphics[scale=0.3]{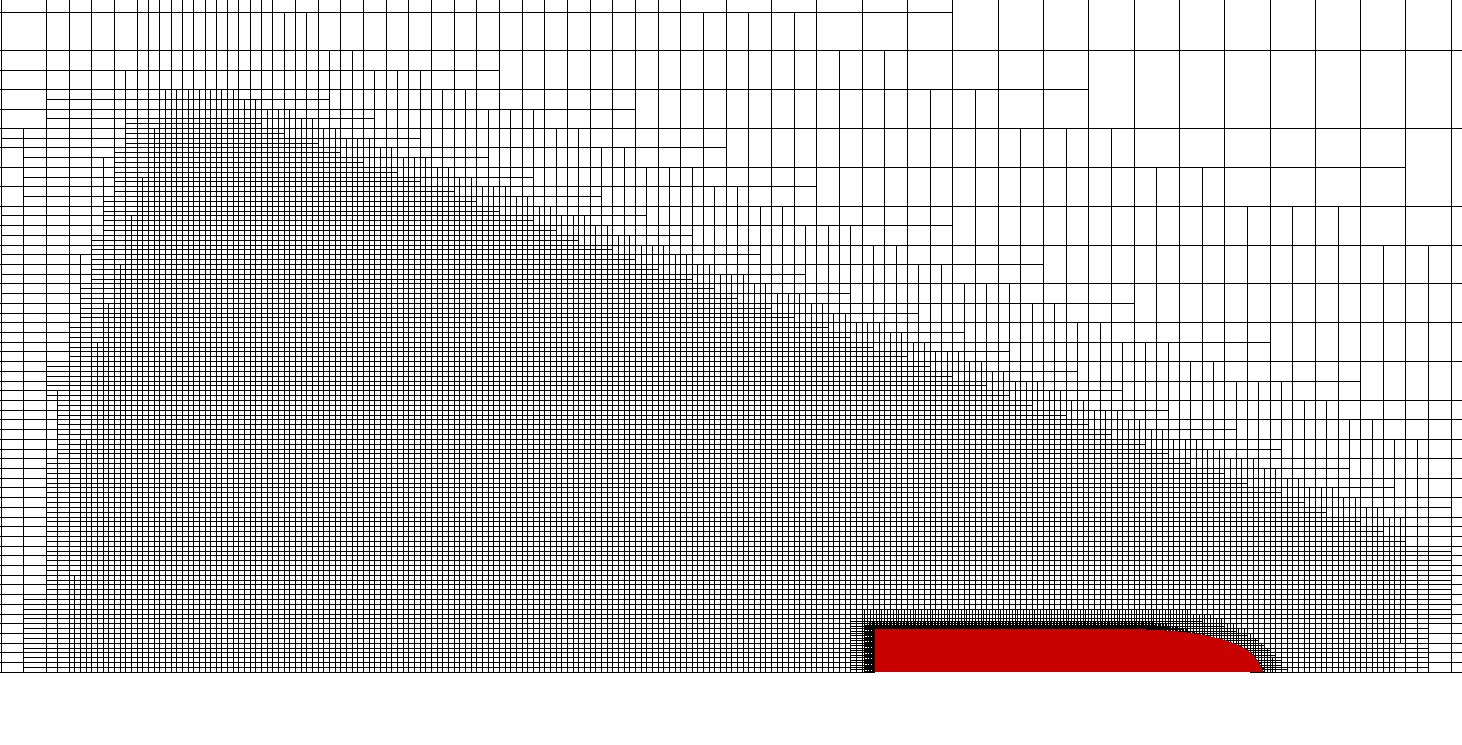}
\caption{Top view on the numerical grid in the calm water plane of the Offshore Supply Vessel.}
\label{fig:osv_fo_grid}
\end{figure}

We define the Reynolds and Froude numbers by reference to the length  $L^\mathrm{ O}$ of the OSV, the cruising speed $v_\mathrm{1}$, the magnitude of the gravity vector $G$ and the kinematic water viscosity $\nu^{\mathrm b}$. The investigated FS configuration yields $\mathrm{Re} = v_\mathrm{1} L^\mathrm{ O}/\nu^{\mathrm b} = \SI{2.81 }{} \cdot 10^8$, $\mathrm{Fn} = v_\mathrm{1}/\sqrt{G L^\mathrm{ O}} = 0.37$. Geometrical constraints considered in the present application refer to (a) the conservation of a plane transom that allows tangential-only deformation, (b) the preservation of the hydrostatic displacement, and (c) the adherence to a maximum length $L^\mathrm{O}$ as well as a maximum permissible hydrostatic draught. The initial OSV consists of a hull, transom, bulkwark, and deck, as conceptually sketched in Fig. \ref{fig:osv_schick}. The analysis is concerned with the steady-state in calm water conditions.
As depicted in Fig. \ref{fig:osv_general}, the origin of the Cartesian coordinate system is located below the transom stern of the initial configuration, and the free surface is initialized in the $x_\mathrm{1}-x_\mathrm{2}$ plane at $x_\mathrm{3} / L^\mathrm{ O} = 1 / 16$.
The simulation domain has a length, height and width of $8 \, L^\mathrm{ O}$, $6 \, L^\mathrm{ O}$ and $4 \, L^\mathrm{ O}$, where the outlet [bottom] boundaries are located four [two] OSV-lengths away from the origin. A dimensionless wave length of $\lambda^\mathrm{FS} = \lambda/ L^\mathrm{ O} = 2 \, \pi \, \mathrm{Fn}^2 = 0.852$ is expected and the total drag of the OSV should be minimized, viz. $r_\mathrm{i} = -\delta_\mathrm{i 1}$ in (\ref{equ:special_objective}). 
The utilized unstructured numerical grid around the transom is displayed in Fig. \ref{fig:osv_general} (b) and consists of approximately $\SI{3}{} \cdot 10^6$ control volumes. Due to symmetry, only half of the geometry is modeled in lateral ($x_\mathrm{2}$) direction. The fully turbulent simulations employ a wall-function-based $k-\omega$ SST model of \cite{menter1994two} together with a non-dimensional wall-normal distance of $y^+ \approx \SI{50}{}$ for the first grid layer adjacent to the hull. The horizontal resolution of the free surface region is refined within a Kelvin-wedge to capture the wave field generated by the vessel, cf. Fig. \ref{fig:osv_fo_grid}. The free surface resolution employs approximately $\Delta x_\mathrm{1} / \lambda = \Delta x_\mathrm{2} / \lambda = 1/50$ cells in the horizontal directions and $\Delta x_\mathrm{3} / \lambda = 1/500$ cells in the vertical direction. Convective primal [adjoint] fluxes are again approximated using the QUICK [QDICK] scheme, cf. \cite{stuck2013adjoint}.
Only the respective approximation of the concentration equation follows again a different approach, cf. Sec. \ref{sec:validation} and \cite{kuhl2021adjoint, kuhl2021cahn}. Simulations are advanced to a steady state in pseudo time at Courant numbers of $\mathrm{Co} \le 0.4$ using an Euler implicit approach. At the inlet, a homogeneous unidirectional (horizontal) bulk flow $v_\mathrm{i} = v_\mathrm{1} \delta_{i 1}$ is  imposed for both phases in conjunction with a calm water concentration distribution. Slip walls are used along the top, bottom, and lateral boundaries, and a hydrostatic pressure boundary is employed along with the outlet. Similar to the validation study, the grid is stretched towards the outlet to suppress the outlet wave field and comply with the outlet condition. The boundary conditions are supplemented by a symmetry condition along the midship plane.

\begin{figure}[!ht]
\centering
\iftoggle{tikzExternal}{
\input{./tikz/osv_primal_initial.tikz}
}{
\includegraphics{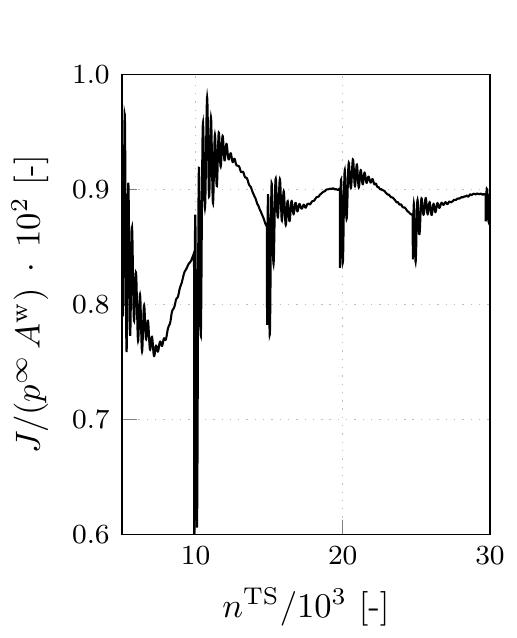}
\includegraphics{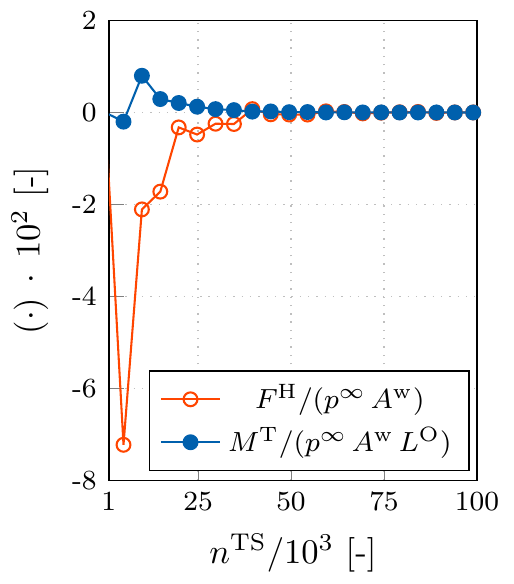}
\includegraphics{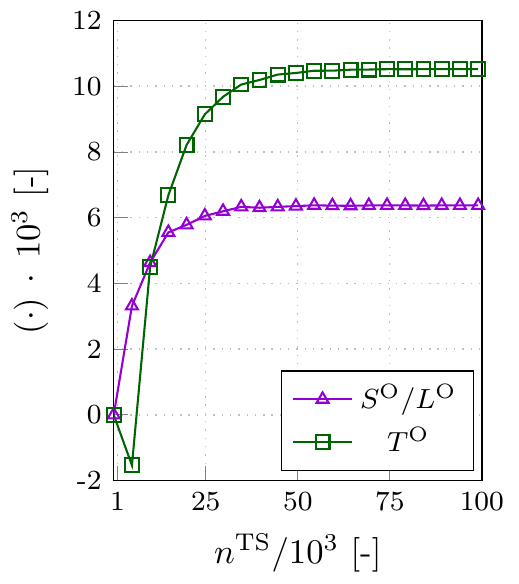}
}
\caption{Dynamic floating position and related forces for the initial OSV at full scale ($\mathrm{Re}_\mathrm{L} = \SI{2.81}{} \cdot 10^8$, $\mathrm{Fn} = 0.37$); (Left) Non-dimensional drag coefficient ($p^\mathrm{\infty} = 0.5 \, \rho^\mathrm{b} \, v_\mathrm{1}^2$), (center) non-dimensional lift and moment coefficient and (right) resulting dimensionless sinkage and trim angle over the number of time steps.}
\label{fig:osv_primal_initial}
\end{figure}

Figure \ref{fig:osv_primal_initial} displays results obtained for the computation of the initial geometry, i.e., the development of the normalized drag (left), normalized heave force $F^\mathrm{H}$ and trim moment $M^\mathrm{T}$ (center), as well as the non-dimensional sinkage $S^\mathrm{O}/L^\mathrm{O}$  and pitch positions $T^\mathrm{O}$ by reference to the initial hydrostatic floatation (right). As indicated by the evolution of the drag, the floatation is adjusted once every 5000-time steps, and the final floating position is found after approximately $n^\mathrm{TS} = \SI{}{40.000}$ steps.
All optimizations allow a tangential deformation of the lateral symmetry plane while maintaining the initial main dimensions of the OSV. Starting from the initial dynamic floating position, two sets of optimizations are performed, each using three different consistency levels for the considered adjoint PDE system. The first triplet neglects the floatation adjustment and requires a smaller number of steps $n^\mathrm{TS}$ to converge the cost functional in pseudo time. The second triplet adjusts the floatation, which results in approximately double the computational effort, cf. Tab. \ref{tab:cpu_time}. The respective three adjoint systems are summarized as follows:
\begin{itemize}
    \item[E1] Experiment 1 neglects both adjoint concentration contributions and adjoint turbulence effects within the adjoint momentum equation, i.e. $\beta = 1$ and $\hat{c} \nabla_\mathrm{i}c \to 0$ in (\ref{equ:adjoint_rans_momentum}). The approach resembles a complete frozen concentration and a frozen turbulence approach, and no need arises to maintain a compressive adjoint concentration transport. The adjoint solution process is iterated in a steady-state manner which drastically reduces the computational time of the adjoint solver.
    \item[E2] The second experiment extends E1 only to improve the influence of adjoint turbulence effects, i.e. $\beta = 2$ and $\hat{c} \nabla_\mathrm{i}c \to 0$ in (\ref{equ:adjoint_rans_momentum}). Therefore, E2 is more consistent w.r.t. adjoint turbulence but still corresponds to a frozen concentration approach, again allowing for a steady adjoint approximation.
    \item[E3] The third experiment maximizes the consistency of the adjoint system  within the scope of this paper. Hence, the adjoint concentration transport is also coupled with the adjoint momentum balance, i.e. $\beta = 2$ and $\hat{c} \nabla_\mathrm{i}c \neq 0$ in (\ref{equ:adjoint_rans_momentum}). Mind that this noticeably increases the computational effort, cf. Tab. \ref{tab:cpu_time}.
\end{itemize}
\begin{table}[!ht]
\begin{center}
\begin{tabular}{|c||c|c|c|}
\hline
Experiment          & E1 [h] & E2 [h] & E3 [h] \\
\hline
\hline
Free Float.         & 19.568    & 19.136    & 33.454 \\
\hline
Fixed Float.        & 10.880    & 12.864    & 26.265 \\
\hline
\hline
Ratio               & 1.798     & 1.487     & 1.273 \\
\hline
\end{tabular}
\end{center}
\caption{Measured computational effort in CPUh ($n^\mathrm{opt} \cdot \overline{t^\mathrm{wc}} \cdot n^\mathrm{CPU}$) for all six optimization studies, where $\overline{t^\mathrm{wc}}$ refers to the mean wall clock time per optimization step and $n^\mathrm{CPU}$ denotes the number of employed CPU cores.}
\label{tab:cpu_time}
\end{table}
To ensure a fair comparison, all optimizations use different step sizes, which are scaled so that each first shape update has a maximum displacement of two per mil of the vessels length, i.e. $d^\mathrm{max} = 2 L^\mathrm{O} / 1000$ in Alg. \ref{alg:optimization_procedure}.

\begin{figure}[!ht]
\centering
\iftoggle{tikzExternal}{
\input{./tikz/osv_optimization_summary.tikz}
}{
\includegraphics{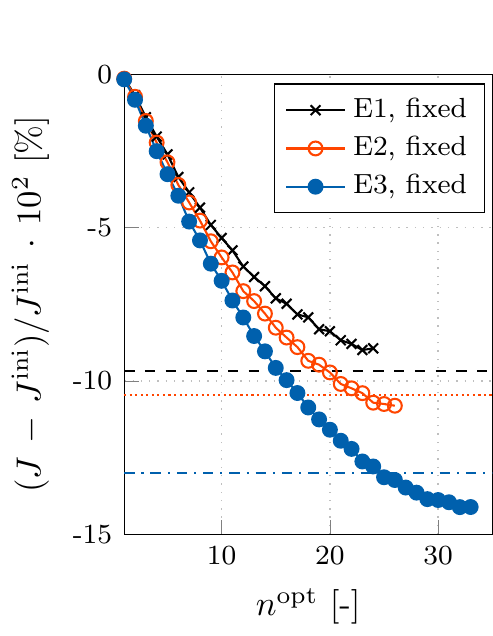}
\includegraphics{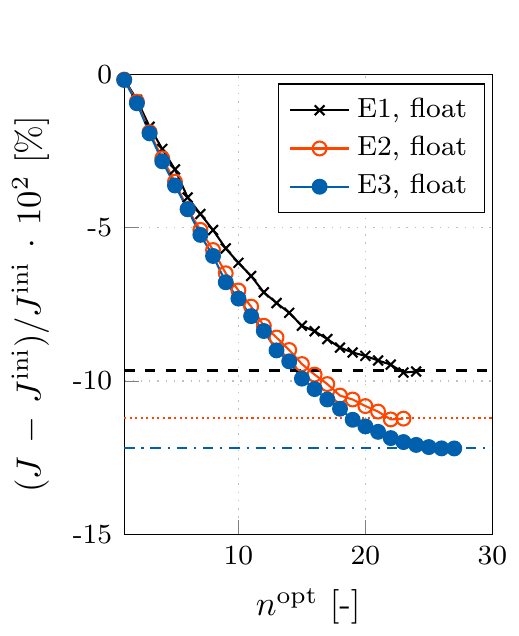}
\includegraphics{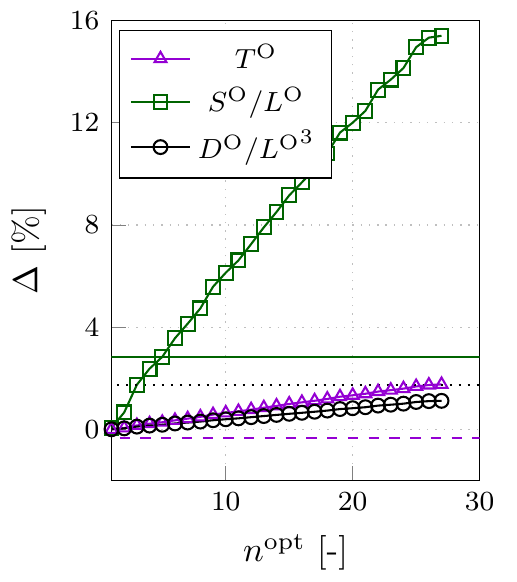}
}
\caption{Drag optimization of an Offshore Supply Vessel (OSV) at full scale ($\mathrm{Re}_\mathrm{L} = \SI{2.81 }{} \cdot 10^8$, $\mathrm{Fn} =0.37$):  Convergences of the optimization for three different adjoint PDE systems (E1-E3) using fixed trim and sinkage (left) and adaptive trim and sinkage (center). Horizontal lines indicate the total resistance obtained with trim and sinkage adjustment after re-simulating the final design from scratch. Right) Relative hydrostatic data for the most consistent (E3) optimization from the center figure.%, where horizontal constant (dashed) [dotted] lines indicate the relative final trim (sinkage) [displacement] variation of the most consistent (E3) and finally released case from the left figure (blue, dashed).
}
\label{fig:osv_optimization_summary}
\end{figure}

Results of the optimizations are shown in Fig. \ref{fig:osv_optimization_summary} for triplet without (left) and with (center) floatation adjustment. All optimizations converge after 20-35 gradient steps. In line with the results of the global validation study reported in Sec. \ref{subsec:global_validation}, a clear trend towards a stronger cost functional decrease $(J - J^\mathrm{ ini}) / J^\mathrm{ ini}$ is observed for the more consistent formulations. While the resistance reduction observed in E1 is single digit ($\approx 9\%$) in both cases, already the algebraic turbulence model (E2) offers an improved drag reduction of about 2-3\% w.r.t. E1. The largest decrease is obtained with the consistent approach E3, which provides an additional 1\% drag reduction for the free floating vessel and even 3\% for the fixed floatation case in comparison to E2. The additional optimization gain justifies the more cost intensive adjoint simulation which refers, to $\mathcal{O}$(\SI{}{20.000}) [$\mathcal{O}$(\SI{}{30.000})] CPUh for the fixed [free] floating E3 case compared to $\mathcal{O}$(\SI{}{10.000}) [$\mathcal{O}$(\SI{}{20.000})] CPUh for the respective E1 scenario, cf. Tab. \ref{tab:cpu_time}.
The E3 optimization with a fixed floating position provides the largest drag reduction of about 13.5\%. Changes of the normalized hydrodynamic floatation for the E3 configuration with floatation adaption are given in the right graph of Fig. \ref{fig:osv_optimization_summary}. The ship trims forward, but sinkage and -- interestingly -- displacement also increase.

After the optimization studies, the respective optimal shapes were re-computed from scratch while being free to adjust their floating position. This effort aims to assess (a) the credibility of the deformation procedure for the triplet that was optimized with an adjustment of the floating position and (b) the uncertainties introduced by neglecting the adjustment of the floating position during a shape optimization. Non-dimensional drag values obtained from these simulations are indicated by the horizontal lines in Fig. \ref{fig:osv_optimization_summary}. As indicated in the center graph, the results of the re-computed optimized geometries are in fair agreement with the final results of the optimization study with an adjustment of the floating position.
As expected, the re-computed drag results deviate from their respective companion results when the floating position is fixed during the optimization. In two cases (E2, E3) the gain decreases and in one case (E1) the result even improves, cf. Fig. \ref{fig:osv_optimization_summary} (left). Though the related drag modifications are limited, reliable predictions only follow optimizations that account for floatation, cf. below. 

\begin{figure}[!ht]
\centering
\iftoggle{tikzExternal}{
\input{./tikz/osv_slices_1.tikz}
}{
\includegraphics{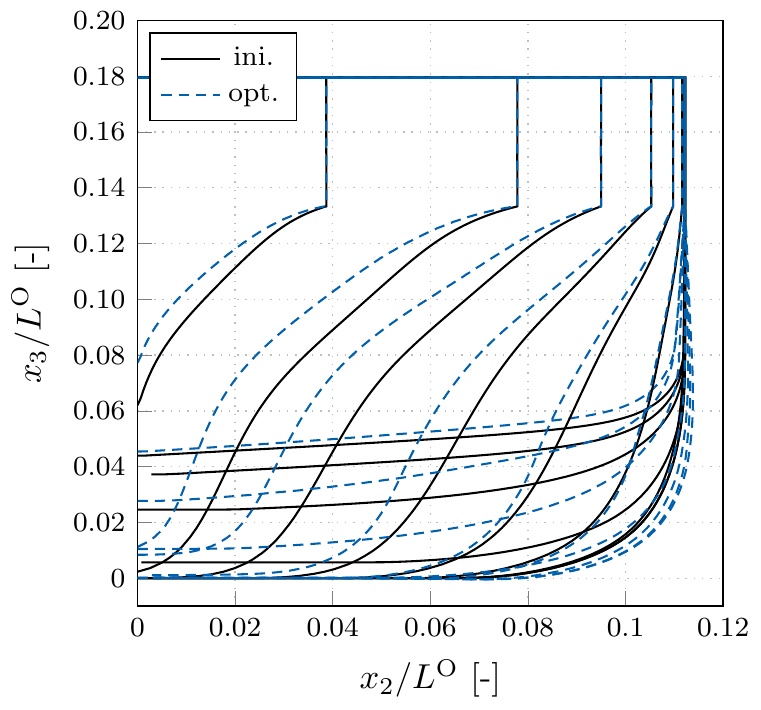}
\includegraphics{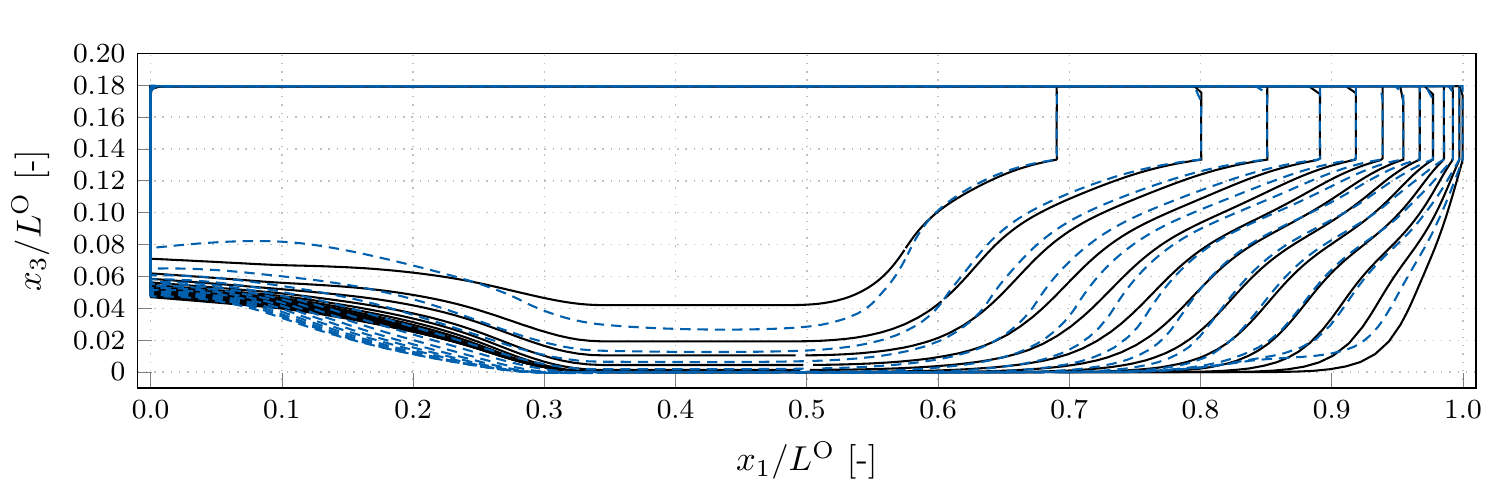}
\includegraphics{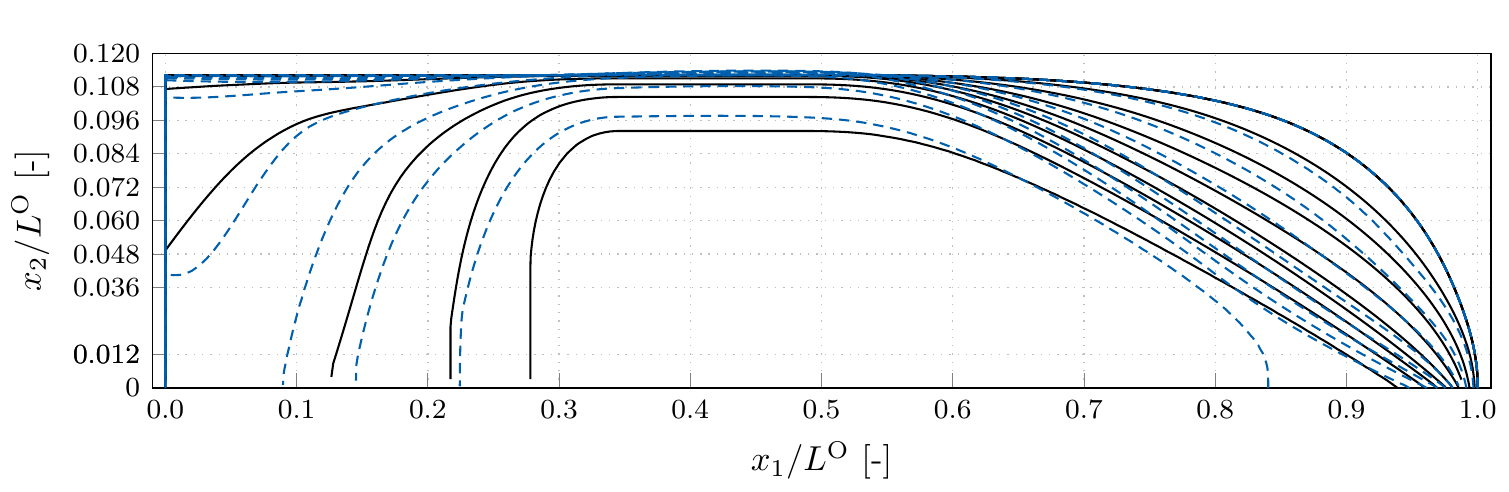}
}
\caption{Drag optimization of an Offshore Supply Vessel (OSV) at full scale ($\mathrm{Re}_\mathrm{L} = \SI{2.81 }{} \cdot 10^8$, $\mathrm{Fn} =0.37$) with adjustment of floating position: Normalized (Top) frames, (center) waterlines and (bottom) buttocks for the initial (black) and optimized (blue, dashed) geometry.}
\label{fig:osv_slices_1}
\end{figure}

\begin{figure}[!ht]
\centering
\iftoggle{tikzExternal}{
\input{./tikz/osv_slices_2.tikz}
}{
\includegraphics{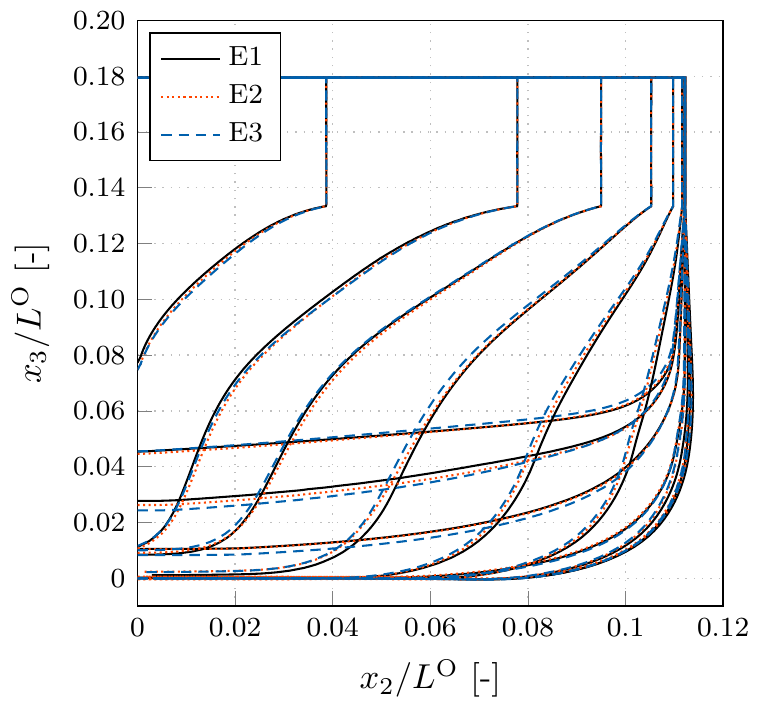}
\includegraphics{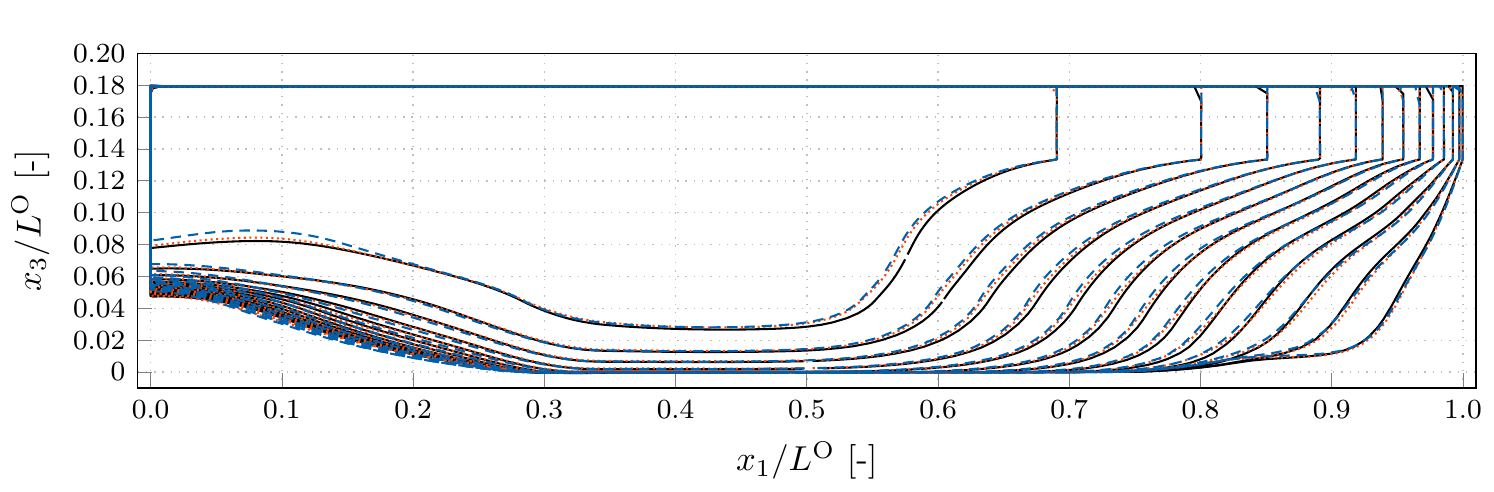}
\includegraphics{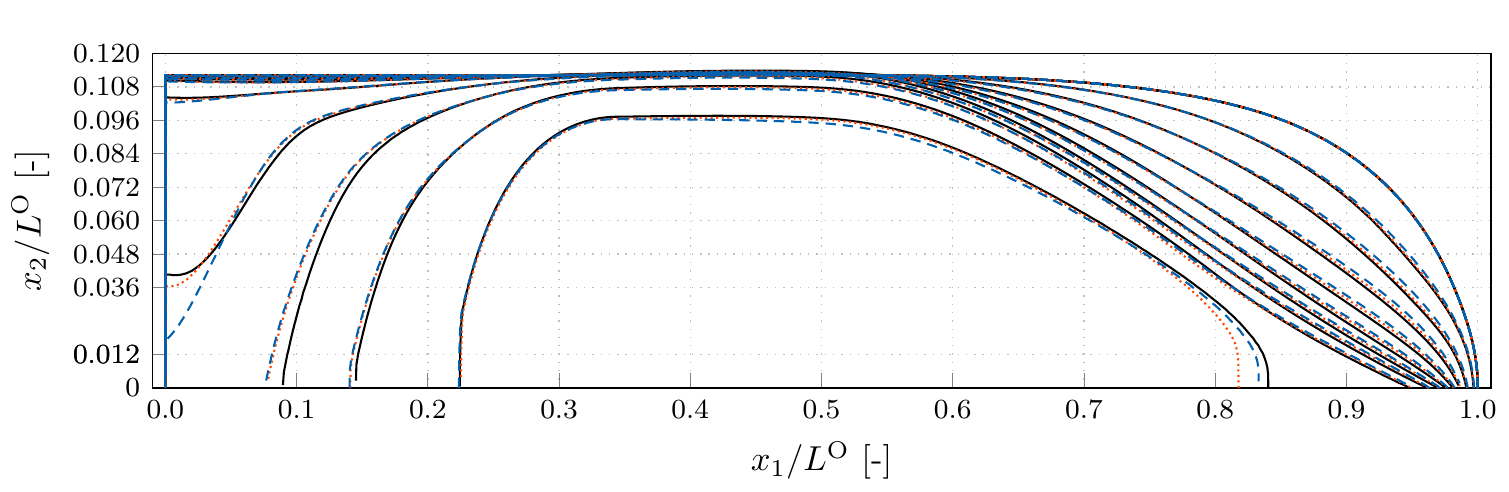}
}
\caption{Drag optimization of an Offshore Supply Vessel (OSV) at full scale ($\mathrm{Re}_\mathrm{L} = \SI{2.81 }{} \cdot 10^8$, $\mathrm{Fn} =0.37$): Comparison of optimized  normalized (Top) frames, (center) waterlines and (bottom) buttocks using different consistency levels, i.e. E1 (black), E2 (orange, dotted), and E3 (blue,dashed).}
\label{fig:osv_slices_2}
\end{figure}

\begin{figure}[!ht]
\centering
\iftoggle{tikzExternal}{
\input{./tikz/osv_slices_3.tikz}
}{
\includegraphics{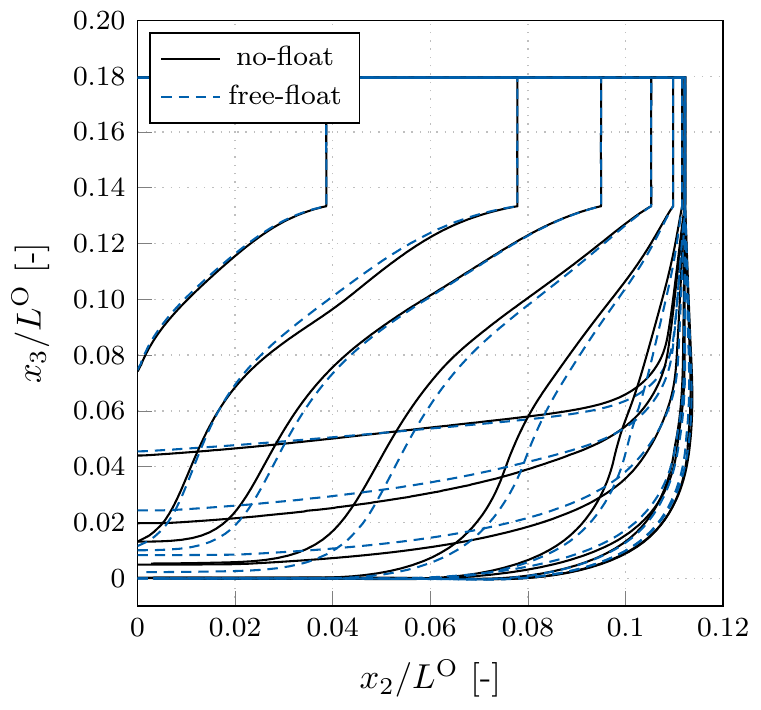}
\includegraphics{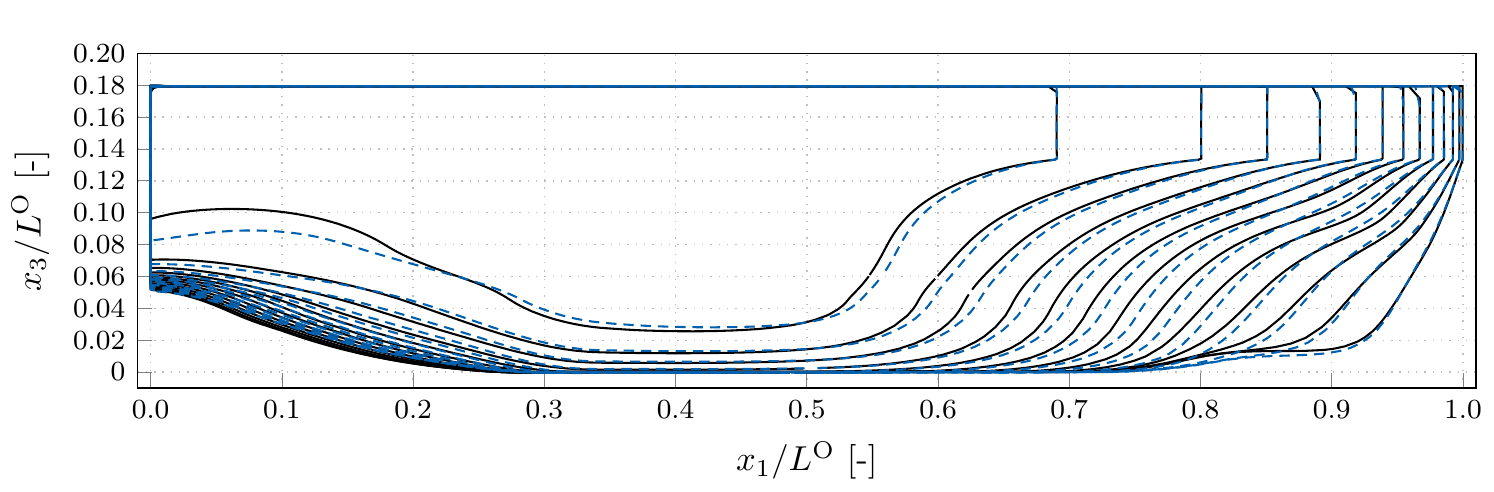}
\includegraphics{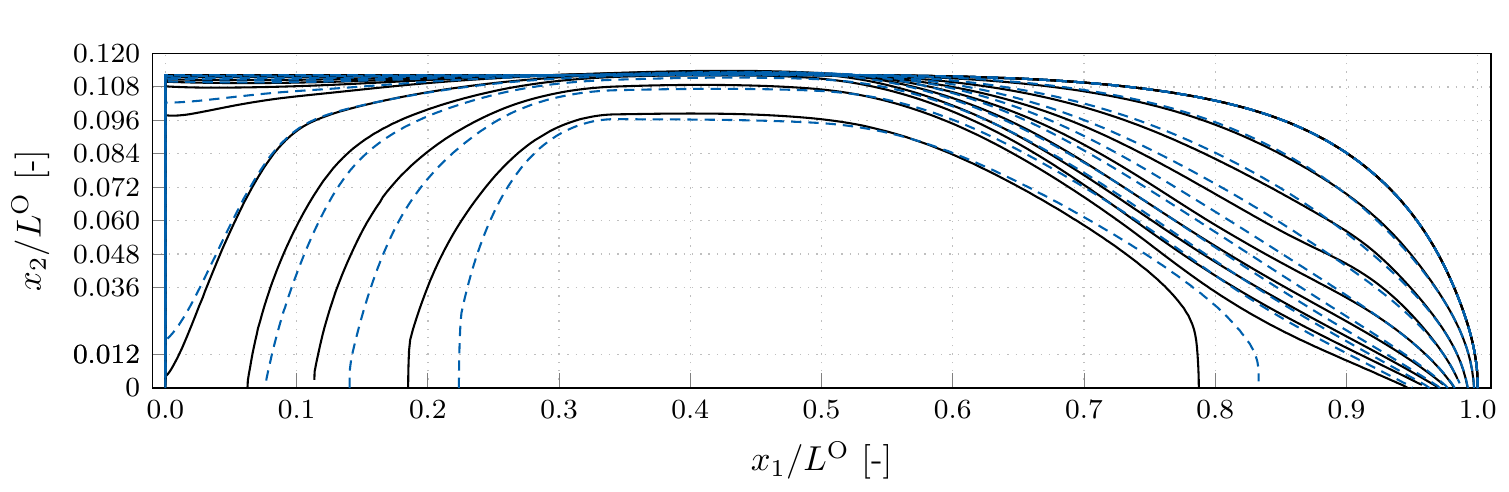}
}
\caption{Drag optimization of an Offshore Supply Vessel (OSV) at full scale ($\mathrm{Re}_\mathrm{L} = \SI{2.81 }{} \cdot 10^8$, $\mathrm{Fn} =0.37$): Normalized (Top) frames, (center) waterlines and (bottom) buttocks for the most consistent (E3) optimized shapes with (blue, dashed) and without (black) floatation adjustment during the optimization.}
\label{fig:osv_slices_3}
\end{figure}

\begin{figure}[!h]
\centering
\iftoggle{tikzExternal}{
\input{./tikz/osv_model_scale.tikz}
}{
\includegraphics{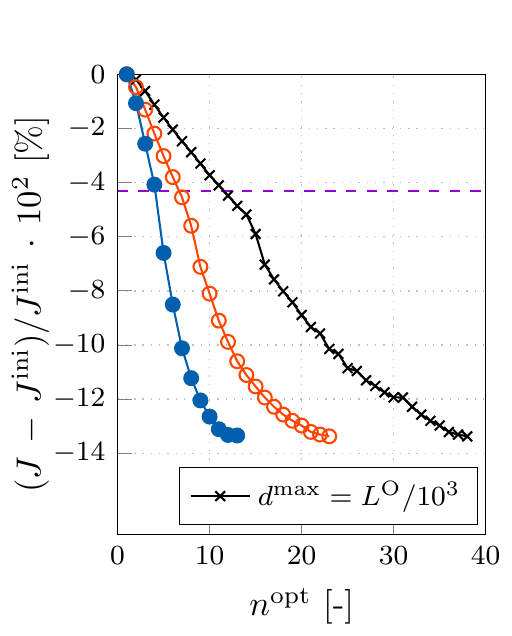}
\includegraphics{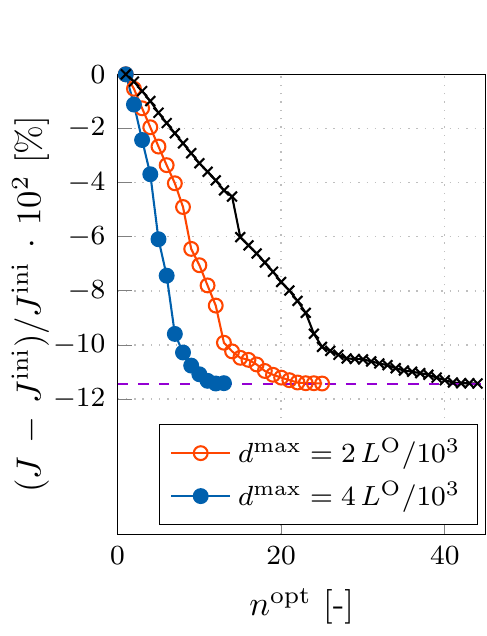}
\includegraphics{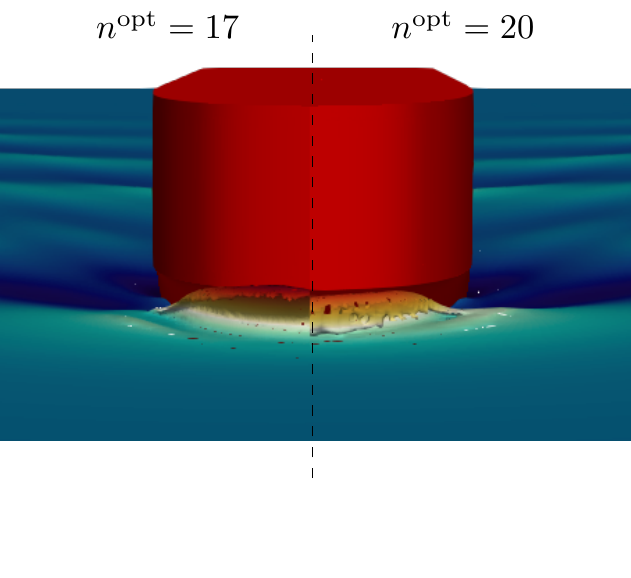}
}
\caption{Drag optimization of an Offshore Supply Vessel (OSV) in model scale ($\mathrm{Re}_\mathrm{L}^\mathrm{MS} = \SI{8.92 }{} \cdot 10^6$, $\mathrm{Fn}^\mathrm{MS} = 0.32$); Convergence of the optimization using fixed (left) and adaptive  (right) floating positions during the optimization. Dashed horizontal lines indicate the  total resistance obtained with adaptive trim and sinkage adjustment  after re-meshing and re-simulating the final designs for $d^\mathrm{max} = L^\mathrm{O} / 1000$ from scratch. Right) Wave field  in the bow region displayed by design candidates 17 (left side) and 20 (right side).}
\label{fig:osv_model_scale}
\end{figure}

Figures \ref{fig:osv_slices_1}-\ref{fig:osv_slices_3} display the optimized hull shapes using frame, water and buttock lines. The observed drag reductions follow from significant changes in the hull shapes. Figure \ref{fig:osv_slices_1} presents frames (top), water lines (middle), and buttocks (bottom) of the initial and the optimized geometry (E3) with the adaption of trim and sinkage. Above all, the S-twist is reduced, the displacement and thickness in the bow area are decreased, and the stern is raised.
A comparison of optimized shapes obtained when adjusting the floatation is shown in Fig. \ref{fig:osv_slices_2} for the three consistency levels E1 (black), E2 (orange, dotted), and E3 (blue, dashed). Qualitatively, the shape changes are similar, but quantitative differences are most pronounced for case E3, particularly in the stern regime. 
Figure \ref{fig:osv_slices_3} compares the optimized geometries obtained from the two E3 configurations (fixed vs. adaptive floatation). The geometries reveal substantial differences regarding the bow and the stern region, with much more pronounced S-shaped outer water lines and a rear-shift of the displacement for the fixed floating position geometry. While this indicates discrepant descent directions, the differences -- notably the surprisingly positive results returned by optimizing for a fixed floating position -- might raise concerns about step size influences.
Therefore a supplementary study on step size influences was performed for a smaller scale configuration, i.e. $\mathrm{Re}^\mathrm{MS} = \SI{8.92 }{} \cdot 10^6$,  $\mathrm{Fn}^\mathrm{MS} = 0.32$. Three different maximum initial deformations, i.e. $d^\mathrm{max} = L^\mathrm{O}/1000$, $d^\mathrm{max} = 2 L^\mathrm{O}/1000$, $d^\mathrm{max} = 4 L^\mathrm{O}/1000$, were investigated in combination with configuration E3. Results of this study are shown in Fig. \ref{fig:osv_model_scale}. The left graph depicts results obtained from a fixed floating optimization and the center graph refers to optimizations with an adjusted floating position. The figure reveals that the step size has no influence on the optimization result. However, in conjunction with a fixed position, the optimization gain drops significantly once the final design is released to find its floating position, cf. horizontal lines in Fig. \ref{fig:osv_model_scale}.
The sensitive interplay between shape modification and floatation can be anticipated from the right graph of the figure. Therein, the free surface, including a breaking bow wave, is compared for two design candidates, i.e., $n^\mathrm{opt}=17$ and $n^\mathrm{opt}=20$.
%to a water elevation from a more optimal hull-shape based on the $c = 0.5$ contour.
A decreased breaking wave amplitude arises during these three optimization steps. This reveals a strong nonlinearity in the design space and significantly influences the floatation -- which has been neglected in this particular study -- and finally yields a substantial trend reversal of the optimization.

%The detection of this nonlinearity, despite the necessity of possibly smaller step sizes, underlines the robustness of the method. Mind that a descent direction along nonlinearities is usually associated with small step sizes

%
As illustrated, the differences experienced from changing the floating position of a hull optimized in a fixed position can be fairly detrimental. Hence, adjusting the floating position is highly recommended to secure the optimization efforts and reduce uncertainties.

%% file: tex/conclusion.tex
The paper reports a node-based continuous adjoint two-phase flow procedure to optimize hull shapes of free-floating vessels. To this end, three topics were addressed that refer to: (1) An adequate two-phase flow model, (2) the relevance of floatation and consistency within the optimization framework, and (3) appropriate descent direction computations that obey local and global technical constraints.

It is seen that elements of a Cahn-Hilliard model should augment frequently employed VoF two-phase flow models to facilitate dual consistency. With attention is restricted to industrial flow simulations that do not resolve the interface physics, i.e. in the discrete sharp interface limit, related modifications are lucid and result in a nonlinear diffusion supplement to the primal/adjoint concentration transport. The authors are convinced that this is crucial for a robust primal/adjoint coupling in marine engineering applications, particularly when attention is devoted to full-scale optimizations at large Reynolds and Froude numbers. 

The paper supports the endeavor for adjoint consistency. An algebraic augmentation of the adjoint eddy viscosity -- which was recently suggested by reference to log-law physics -- returns noticeable benefits with around 2\% increased drag reduction for the optimal configuration. More importantly, the concentration contribution to the adjoint momentum equation should be considered to expose the full potential of the adjoint optimization. Furthermore, all density variation terms of the adjoint concentration equation must be considered to secure a gradient descent.     
 
Load variations -- induced by the shape update -- alter the vessel's floatation, particularly the trim and sinkage, which in turn yield modified loads. To mitigate the related uncertainties, adjusting the floating position during the optimization is highly recommended. The present study suggests that a ''frozen adjoint floatation'' approach is sufficient for steady-state resistance optimizations.    

The Steklov-Poincaré metric offers an efficient gradient computation strategy that intensively re-uses the established coding infrastructure of a CFD algorithm. Its merits refer to the simultaneous update of the volume and the surface mesh of the optimized shape. Furthermore, the procedure can easily be customized to obey local and global technical constraints on the vessel's displacement, extensions, and further design demands like, e.g., a plane transom.